\documentclass[manuscript,nonacm]{acmart}
\usepackage{wrapfig}
\usepackage{subcaption}
\usepackage{multirow}

\usepackage{amsmath}
\usepackage{subcaption}

\usepackage{changepage}
\usepackage{array}
\usepackage{textcomp,marvosym}

\usepackage{nameref,hyperref}
\AtBeginDocument{%
  \providecommand\BibTeX{{%
    \normalfont B\kern-0.5em{\scshape i\kern-0.25em b}\kern-0.8em\TeX}}}

\title{Machine Learning-based Context Aware EMAs: An Offline Feasibility Study}

\author{Zachary D King}
\affiliation{%
  \institution{Department of Electrical and Computer Engineering, Rice University}
  \city{Houston}
  \state{TX}
  \country{USA}}

\author{Maryam Khalid}
\affiliation{%
  \institution{Department of Electrical and Computer Engineering, Rice University}
  \city{Houston}
  \state{TX}
  \country{USA}}

\author{Han Yu}
\affiliation{%
  \institution{Department of Electrical and Computer Engineering, Rice University}
  \city{Houston}
  \state{TX}
  \country{USA}}

\author{Kei Shibuya}
\affiliation{%
  \institution{Biometrics Research Laboratories, NEC Corporation}
  \city{Tokyo}
  \country{Japan}}

\author{Khadija Zanna}
\affiliation{%
  \institution{Department of Electrical and Computer Engineering, Rice University}
  \city{Houston}
  \state{TX}
  \country{USA}}

\author{Marzieh Majd}
\affiliation{%
  \institution{Department of Psychiatry, Brigham and Women’s Hospital}
  \city{Boston}
  \state{MA}
  \country{USA}}
\affiliation{%
  \institution{Department of Psychiatry, Harvard Medical School}
  \city{Boston}
  \state{MA}
  \country{USA}}

\author{Ryan L Brown}
\affiliation{%
  \institution{Department of Human Development and Family Sciences, Texas Tech University}
  \city{Lubbock}
  \state{TX}
  \country{USA}}

\author{Yufei Shen}
\affiliation{%
  \institution{Department of Electrical and Computer Engineering, University of Texas}
  \city{Austin}
  \state{TX}
  \country{USA}}

\author{George Kypriotakis}
\affiliation{%
  \institution{Department of Behavioral Science, University of Texas MD Anderson Cancer Center}
  \city{Houston}
  \state{TX}
  \country{USA}}

\author{Thomas Vaessen}
\affiliation{%
  \institution{Department of Behavioural, Management and Social Sciences, University of Twente}
  \city{Enschede}
  \country{Netherlands}}

\author{Christopher P Fagundes}
\affiliation{%
  \institution{Department of Psychological Sciences, Rice University}
  \city{Houston}
  \state{TX}
  \country{USA}}
  \affiliation{%
  \institution{Institute for Health Resilience \& Innovation, Rice University}
  \city{Houston}
  \state{TX}
  \country{USA}}

\author{Akane Sano}
\affiliation{%
  \institution{Department of Electrical and Computer Engineering, Rice University}
  \city{Houston}
  \state{TX}
  \country{USA}}

\begin{document}

\renewcommand{\shortauthors}{King et al.}
\begin{abstract}
Mobile health (mHealth) systems help researchers monitor and care for patients in real-world settings. Many studies that utilize mHealth applications use Ecological Momentary Assessments (EMAs), passive sensing, and contextual features to develop emotion recognition models, which rely on EMA responses as ground truth. Due to this, it is crucial to consider EMA compliance when conducting a successful mHealth study. Utilizing machine learning is one approach that can solve this problem by sending EMAs based on the predicted likelihood of a response. However, literature suggests that this approach may lead to prompting participants more frequently during emotions associated with responsiveness, thereby narrowing the range of emotions collected.

We propose a multi-objective function that utilizes machine learning to identify optimal times for sending EMAs, thereby increasing the likelihood of response and capturing a broader range of emotions. The function identifies optimal moments by combining predicted response likelihood with model uncertainty in emotion predictions. Uncertainty would lead the function to prioritize time points when the model is less confident, which often corresponds to underrepresented emotions. We demonstrate that using this objective function to guide EMA delivery would result in prompts being sent when participants are responsive and experiencing less commonly observed emotions. Prioritizing less commonly observed emotions promotes emotional diversity, defined as the range of emotions a person experiences over time, leading to a more comprehensive picture of overall well-being.

The evaluation is conducted offline using two datasets: (1)  91 spousal caregivers of individuals with Alzheimer’s Disease and Related Dementias (ADRD), (2) 45 healthy participants. Results show that the multi-objective function tends to be higher when participants respond to EMAs and report less commonly observed emotions. This suggests that using the proposed objective function to guide EMA delivery could improve receptivity rates and capture a broader range of emotions. 
\end{abstract}



\maketitle

\section{Introduction}

Mobile health (mHealth) applications and systems have seen a remarkable rise in popularity. mHealth research has demonstrated significant promise in aiding healthcare providers in managing various mental health conditions, offering cost-effective, easily accessible, and minimally intrusive interventions \cite{rowland2020clinical}. The primary advantage of these mHealth systems lies in their capacity to gather data (through wearable or phone sensors) that can help predict adverse health outcomes. This, in turn, enables researchers to develop interventions through reminders, education, or motivation \cite{luxton2011mhealth} in real-time. 

Ecological Momentary Assessments (EMA) capture daily in-situ snapshots of individuals' emotions and behaviors through recurring surveys \cite{schwartz1998strategies} and are commonly used in mHealth applications. EMAs are particularly important in mental health and well-being applications and serve as the gold standard for measuring emotional state. Without EMA responses, mHealth applications will struggle to accurately assess users' emotional states. However, EMAs can be burdensome to users as they require active responses, especially when prompts are frequent. Rather than prompting users more often, mHealth studies can collect more responses by improving participant compliance.

Some researchers have utilized machine learning techniques to determine the most appropriate time to send an EMA or intervention, which has proven effective in improving receptivity rates for both EMAs \cite{mishra2017investigating} and interventions \cite{mishra2021detecting}. Several researchers have explored alternative mechanisms to enhance EMA receptivity, including reducing the complexity or frequency of prompts or increasing incentives \cite{eisele2020effects, wen2017compliance, intille2016muema}. Yet, these approaches may limit the number of responses, alter survey content, or increase study costs. Conversely, machine learning-based interruptions can maintain higher compliance rates without these drawbacks. 

However, a significant challenge in using machine learning to schedule EMAs is its potential unintended consequences on reported responses. Several researchers have explored factors influencing interruptibility, such as respondents' emotional state \cite{ho2005using}. Specifically, there is a well-documented link between emotional state and EMA responsiveness \cite{murray2023prompt, rintala2020momentary, wen2017compliance}, with several studies indicating that participants are less likely to respond when experiencing negative emotions. Emotional state and several other factors can be used as a proxy for receptivity; however, these factors would affect the prediction of a receptivity model and lead to the model avoiding prompts when participants are experiencing emotions associated with non-responses \cite{king2024investigating}. This poses a challenge for emotion recognition models. If machine learning-based EMA triggers disproportionately send notifications during certain emotional states, it may inadvertently hinder the development of an emotion recognition model.

\begin{figure}
\centering  
\includegraphics[width=0.98\linewidth]{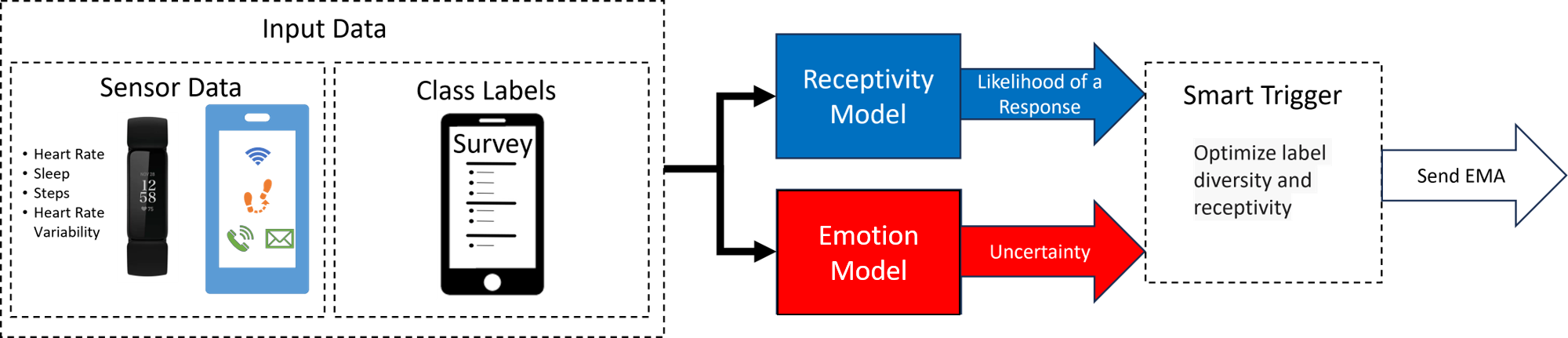}

\caption{Framework for an EMA Trigger that would utilize the proposed multi-objective function.}
\label{fig:frame}
\end{figure}

To address this challenge, we propose a multi-objective function designed to optimize the timing of EMAs using data from wearable devices and phone sensors. This function incorporates two machine learning models: one estimates the likelihood of an EMA response, and the other captures uncertainty in predicting the study-specific emotional state. A trigger would then deliver the EMA at the time that maximizes the objective function’s output. The potential framework of this Smart Trigger is shown in Figure \ref{fig:frame}. This approach balances both the likelihood of participant response (EMA receptivity) and the uncertainty in predicting the target emotion. The inclusion of the receptivity model in the multi-objective function aims to improve the participant’s receptivity rate. While prompting during periods of high model uncertainty encourages EMA delivery across a broader range of emotional states, as uncertainty tends to be higher for emotions that are less frequently represented in the training data. Accounting for uncertainty when sending EMAs helps counterbalance any potential biases the receptivity model may have on reported emotions.  As stated, a system that prompts users based solely on their likelihood of responding would send EMAs more frequently during emotions associated with responsiveness. In contrast, using the proposed objective function would minimize this effect.

The main objective of this work is to demonstrate the feasibility of our multi-objective function in two distinct populations. Insights gained from this work will inform the development of a future ML-based context aware EMA and the design of a future validation study. A significant gap in current research lies in the lack of adaptive methods for improving EMA receptivity rates. Existing machine learning-based EMA triggers primarily focus on responsiveness as the sole criterion for delivering EMAs, often overlooking the collection of responses during varying emotions. This work addresses this gap by introducing a multi-objective function designed to deliver EMAs during moments when participants are both responsive and when the emotion recognition model is uncertain. To achieve this, we address three key research questions (\textbf{RQs}): 

\begin{itemize}
  \item[] \textbf{RQ1:} \textit{Would the proposed multi-objective function send EMAs when people are more responsive?}
  \item[] \textbf{RQ2:} \textit{Would the proposed multi-objective function send EMAs when participants are experiencing emotional states that are less observed?}
  \item[] \textbf{RQ3:} \textit{To what extent would the multi-objective function improve participant compliance and enhance the distribution of captured emotional responses compared to traditional EMA delivery methods?}
\end{itemize}

\section{Related Work}

\subsection{Mobile and Sensor-based Emotion Recognition}

Momentary emotion recognition is a component in mHealth research, enabling timely interventions that support mental health and complement traditional care approaches \cite{jameel2022mhealth}. To predict momentary emotions, researchers use a wide range of sensors, signals, and features including, respiration \cite{bari2020automated, he2017emotion, huynh2021stressnas}, electrocardiogram (ECG) \cite{yu2023semi, hu2018scai, sano2013stress}, Galvanic Skin Response (GSR) \cite{zhao2018emotionsense}, skin temperature \cite{huynh2021stressnas}, accelerometers \cite{rashid2020predicting}, and phone-based data including Global Positioning System (GPS) and call logs \cite{nalepa2019analysis}. The ability to predict these constructs varies based on the sensor suite, the complexity of the intended construct, and the algorithm. The significant pitfall of these momentary emotion recognition models is their dependency on EMA responses as ground truth. As sensors generate increasingly rich data, sparse EMA responses continue to be a bottleneck. Addressing this mismatch will require innovations in labeling methods and improvements in EMA design or delivery to support the development of more effective and scalable emotion-aware systems.

\subsection{EMA}
As mentioned, many researchers have studied the factors influencing receptivity or interuptability. Ho et al. described 11 components that affect the perceived burden of an interruption, including present and future activity, frequency and modality of the interruption, the utility of the message, the required effort, the interface, social aspects, history of responding, and the user's emotional state \cite{ho2005using}. Some of the more common solutions proposed to improve receptivity and limit the burden of an interruption that we found in the literature include changing the sampling rate (frequency), reducing the length of the EMA (complexity), and increasing the compensation of the study or increasing compensation per response (incentivization).

Researchers often explore adjusting EMA frequency or complexity to improve compliance, but the effects on receptivity remain mixed. Some studies, including Eisele et al. and Field et al., found little to no relationship between EMA frequency and perceived burden \cite{eisele2020effects, jones2019compliance}, while others, such as Wen et al., reported higher compliance when EMA frequency is reduced among Youth in non-clinical studies \cite{wen2017compliance}. Despite these findings, reducing EMA frequency limits data density, posing challenges for machine learning applications. An alternative strategy employed by several researchers is to reduce the size or complexity of the EMA. Eisele et al. showed that smaller question sets reduce perceived burden \cite{eisele2020effects}. Intille et al. introduced the $\mu$-EMA, which prompted users more frequently with fewer questions and maintained stable compliance over four weeks, unlike traditional EMAs, which saw declines \cite{intille2016muema}. Li et al. proposed a similar method but used a dynamic prompt that only provided questions based on the potential information gained \cite{li2024ask} from each question. Still, reducing or omitting items from validated instruments may compromise construct validity and limit the interpretability of results. Furthermore, EMAs with fewer questions also provide less information and context, which may reduce the overall value of the collected data.

Incentives are another strategy used to improve EMA receptivity. Wrzus et al. found that 15\% of EMA studies offered no incentives, while 27\% tied incentives to compliance \cite{wrzus2023ecological}. Ottenstein et al. found that EMA studies offering incentives tend to achieve higher compliance rates than those that do not \cite{ottenstein2022compliance}. Although effective for recruitment, incentive-based approaches raise ethical concerns about coercion \cite{vellinga2020patients} and may be unfair if participants miss prompts due to scheduling conflicts. A more sustainable approach would focus on improving compliance without relying on compensation, especially for applications intended for the consumer market, where users engage with mHealth applications for their own well-being.

A more recent method employed by researchers involves utilizing sensors and contextual data to help determine when to send EMAs. Studies have shown that factors like location, personality traits, physical activity, and time of day \cite{morrison2017effect, pielot2017beyond, bidargaddi2018prompt, kunzler2019exploring, mehrotra2015designing} play a role in receptivity.  Mishra et al. proposed using machine learning to improve the receptivity of Just-in-time interventions (JITI) \cite{mishra2021detecting}. They tested a static model using only the previously collected data and an adaptive model, which expanded on the static model using information gathered from the user. Similar to the previous methods discussed, this method has drawbacks. The primary issue with an ML-based EMA trigger is mentioned by King et al., who detailed the relationship between emotional state and EMA receptivity \cite{king2024investigating}. Their findings suggest that sending EMAs based on the predicted likelihood of a response would introduce bias into participant responses. This bias is that notifications would be sent during more positive emotions compared to if the EMAs were sent randomly. However, this method avoids the drawbacks of the previous three methods. 


\section{Multi-Objective Function for Optimal EMA Delivery}

The proposed multi-objective function combines two models: a receptivity model that predicts the probability of an EMA response and an emotion recognition model that provides the uncertainty of an emotion prediction at any given moment. The objective function integrates the outputs of both models to deliver the EMA at a time that maximizes the combined likelihood of response and the uncertainty of the emotion recognition model. This could improve receptivity rates and emotion prediction accuracy by targeting moments of high model uncertainty. This strategy also helps the model to personalize and converge more rapidly than if EMAs were sent randomly. An example of when model uncertainty may be higher is when the model encounters less familiar or previously unobserved patterns in the data. These moments may reflect atypical behaviors, such as attending a unique event, exercising more than usual, or experiencing an emotion that has not been reported before. 

This section includes the methods used for model development and evaluation. In the data processing subsection, we discuss feature processing and extraction, timezone drift correction, and class labels. The model development subsection describes the design of the emotion recognition and the receptivity models. We then introduce the proposed multi-objective function. And finally, the Evaluation subsection discusses our methods for evaluating model performance and the multi-objective function.

\subsection{Receptivity and Emotion Models}

The receptivity model is trained and tested using binary receptivity labels, whereas the emotional state model is trained and tested on sequential labels representing emotion. The input features for these two models are computed from sensor signals. These models operate on segments of data that are 30 minutes long. Each segment is assigned a label: non-responsive, responsive, or unlabeled. An emotion score is also assigned to that segment if there was a response. More information on the reported emotional states can be found in section ~\nameref{dataset_eval}.

\subsubsection{Class Labels}

\begin{figure}[!ht]
\includegraphics[width=0.9\linewidth]{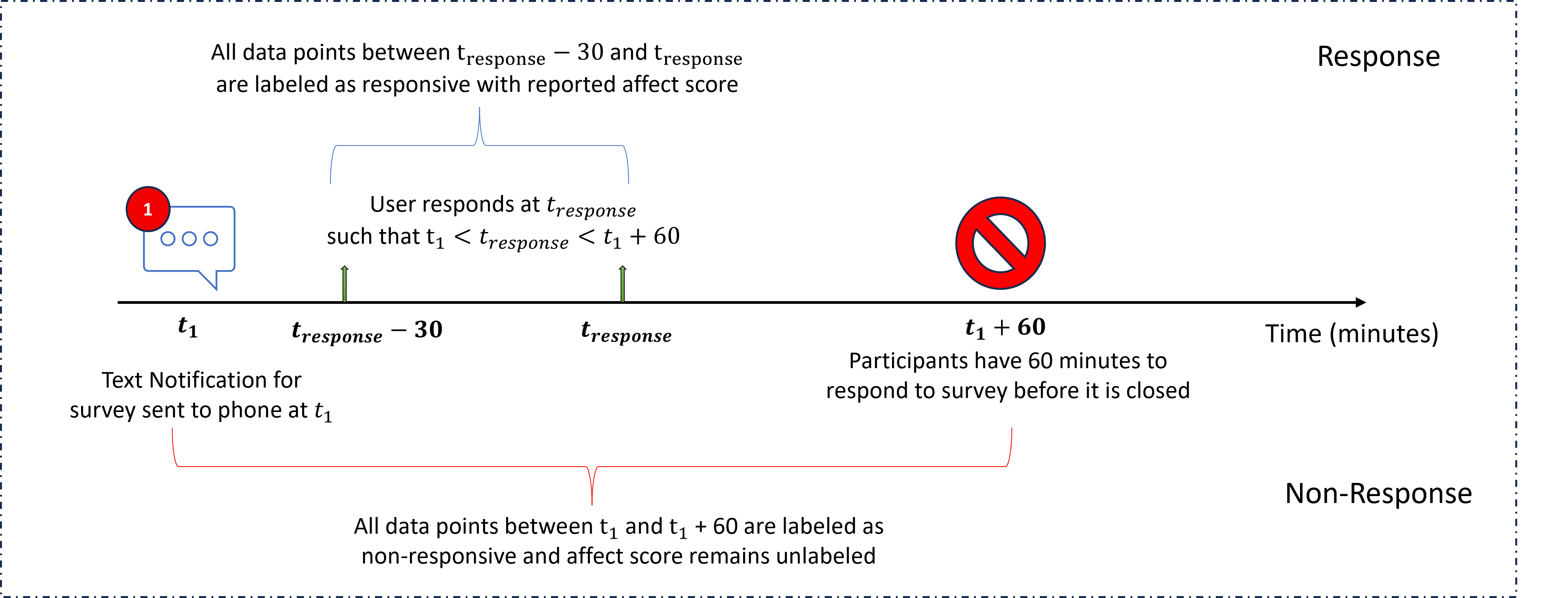}
\caption{Receptivity Labeling Methodology for Responses (top) and Non-Responses (bottom). In the figure, $t_1$ refers to the time a notification was sent, and $t_{response}$ refers to a participant's response time.}
\label{fig:receptivity-label}
\end{figure}

Receptivity was labeled based on the time of the initial notification of the EMA and when the user responded, as shown in Fig. \ref{fig:receptivity-label}.  If a person did not respond to the EMA, the entire 60 minutes after the time of notification was labeled as non-receptive. Since participants have 60 minutes to respond to the survey before it closes. If the participant responds to the survey, the 30 minutes before the response are labeled as receptive. We label the prior 30 minutes as receptive rather than the entire 60 minutes because we cannot assume that the participant will remain receptive after responding to an EMA.  We chose a 30-minute window to increase the number of labeled data points. Further increasing the window size would make it more challenging to argue that the labeled data points are associated with receptivity and emotional state. However, in the Appendix (Tables ~\ref{tab:S3} and ~\ref{tab:S5}, we present the performance metrics of both the receptivity and emotion recognition models using varying window sizes (10, 15, 20, and 60 minutes). After comparing different window sizes, we found that the 30-minute window consistently yielded the best performance for both datasets in predicting emotional state and receptivity. Like labeling receptivity, the segments within the 30-minute window are assigned emotion scores based on the response. 

\subsubsection{Receptivity Model}

The class labels for the receptivity model are binary, indicating whether a response was collected from a prompted EMA or not. We test several machine-learning algorithms selected based on past research for both studies, including Support Vector Machine\cite{kunzler2019exploring, sarker2014assessing}, Random Forest \cite{mehrotra2015designing, kunzler2019exploring, mishra2021detecting, pejovic2014interruptme}, Boosting Algorithms \cite{mehrotra2015designing, pejovic2014interruptme}, Naive Bayes \cite{mehrotra2015designing, pejovic2014interruptme}, and Neural Network \cite{chen2023you, nagesh2021transformers}.  These prior works predict receptivity; however, some focus on EMA responsiveness, while others target JITAI responsiveness. In addition to these models, we employ a simple neural network with two layers and output dimensions of 16 and 8. Both layers use a ReLU activation function, while the output layer applies a sigmoid activation function. In addition, we include a random baseline model that predicts responsiveness based on the user's receptivity rates.

\subsubsection{Emotion Model}
Our emotion model's class labels are ordinal; therefore, we treat the emotion model as a regression problem. Unlike the receptivity model, our multi-objective function uses model uncertainty rather than the prediction. Obtaining uncertainty measures for a regression model is not as straightforward as for categorical or binary labels. For this reason, we designed a neural network that utilizes dropout to capture model uncertainty.  The output dimensions of each layer are 64, 32, and 16, and use a Leaky-ReLU activation function. The loss function employed in this model is a mean square error (MSE). During training, we apply a dropout layer (dropout rate = 0.3) after the first two activation layers. And during testing, we perform 200 stochastic forward passes with dropout enabled, allowing us to generate a distribution of predictions for each input and estimate model uncertainty. The mean of these predictions is used as the final emotion score estimate, and the variance across predictions is used to measure model uncertainty.

In addition to modeling emotion using the neural network, we also model emotion using linear regression and a random baseline. The baseline model for emotional state is chosen randomly based on the mean and standard deviation of emotion scores in the training set. These models are used only for performance evaluation, as the proposed objective function relies on uncertainty for decision-making. Most emotion recognition studies use classification models, simplifying continuous emotion into categories (e.g., low, medium, high). Prior work in regression-based emotion recognition often uses inputs like speech or images to predict valence and arousal \cite{akhand2021facial, galvao2021predicting, mitenkova2019valence}. Our approach uses regression to generate uncertainty estimates that highlight underrepresented emotional states. This avoids sacrificing the nuance of our labels that occurs when categorizing emotion scores.


\subsubsection{Multi-Objective Function}

The multi-objective function helps determine the optimal time for an EMA based on the output of the receptivity model (likelihood for a response) and the uncertainty of the predicted emotional states. This is similar to the function presented by Kuo et al. \cite{kuo2018cost}. Equation \ref{opt} represents the multi-objective function. The $w_u$ and $w_r$ variables are weights that allow the researchers to prioritize either uncertainty or receptivity. The functions $U(t)$ and $R(t)$ are computed from the receptivity and emotion model outputs at time $t$, where $t \in T$ represents the time frame for sending an EMA (ex., between 9 AM and 12 PM). For receptivity, $R(t)$ is the probabilistic prediction of a response, and $U(t)$ represents the uncertainty measure. Uncertainty and receptivity are maximized with respect to $t$, corresponding to the ideal time to send an EMA.

\begin{equation} \label{opt}
 J = \max_{t \in T} \; w_u U(t)^2  + w_r R(t)^2
\end{equation}


\subsection{Datasets}\label{dataset_eval}

We evaluate the multi-objective function using data from two studies, each of which collected wearable and EMA data. One study involved ADRD spousal caregivers (hereafter referred to as the 'ADRD study'), while the other involved healthy participants (hereafter referred to as the 'Healthy study'). 

\subsubsection{Ethical Considerations}
The ADRD study (IRB-FY2021-65) has been reviewed and approved by the Rice University Institutional Review Board. The Healthy study was granted ethical approval by the Sociaal-Maatschappelijke Etische Commissie of Katholieke Universiteit Leuven (G-201809 1339).

\subsubsection{ADRD Study}

\textbf{Participants and Study Design:}
The ADRD study aimed to understand the mental and physical effects associated with ADRD spousal and family caregivers (N=91). We included 73 participants in our analysis after removing participants with no EMA responses overlapping with sensor data (n=14) and those who dropped out (n=4). Excluding the 4 participants who dropped out, the average age was 62 (SD = 11.8), and 87\% were female. The gender disparity reflects that women make up about two-thirds of spousal caregivers for Alzheimer’s patients \cite{alzheimer20192019}. The remaining demographic information for the ADRD dataset is found in the appendix (Table~\ref{tab:S1}). The ADRD study included three blocks of data collection, each lasting one week. During each data collection block, participants were asked to respond to 5 EMAs daily (1 morning, 1 bedtime, and 3 random), wear a Fitbit Inspire 2, and download the AWARE application \cite{ferreira2015aware}. 

\textbf{EMA:}
Each EMA was randomly sent within one of five scheduled time blocks spaced throughout the day via text and email. Each message contained a link to a Qualtrics survey. The survey was available for 1 hour, and participants received a reminder text after 30 minutes. The morning EMA was sent between 8 and 9 AM and included questions about the quality and length of their sleep. The bedtime EMA was sent between 8 and 9 PM and included questions reflecting the day. The remaining three EMAs were composed of the Positive and Negative Affect Schedule (PANAS) \cite{mackinnon1999short}, along with questions regarding depression and loneliness. We use the shortened 10-item PANAS questionnaire, consisting of five positive affect (PA) and five negative affect (NA) items rated on a scale from 1 to 5. PA and NA composite scores are calculated by summing the respective items, resulting in scores ranging from 5 to 25.

\textbf{Sensor Data and Processing:}
Using Fitbit's intra-day API, we extracted minute-by-minute data for steps, heart rate, sleep, and Heart Rate Variability (HRV). We extracted sleep duration, efficiency, and regularity\cite{phillips2017irregular}. HRV features were calculated during a 5-minute window and included the root mean sum of square difference (RMSSD), High Frequency (HF), and Low Frequency (LF). HRV data was collected exclusively during sleep periods lasting more than three hours when the participant was relatively still. 

The AWARE app collects four data modalities: phone logs, message logs, screen usage, and location. From these, we calculate several features, including clustered location, screen time, call duration, number of calls (incoming, outgoing, and missed), and number of SMS messages (incoming and outgoing). The data was temporarily stored on the phone and then sent to a MySQL server at Rice University when the participant's phone was charging and connected to WiFi. A list of features extracted can be found in Table ~\ref{tab:S6}, and a more detailed description of data processing and feature extraction can be found in the appendix (section \ref{Dataset_App}).


\subsubsection{Healthy Study}

\textbf{Participants and Study Design:}
Forty-five healthy participants enrolled in the study to assess the impact of laboratory stress tests on participants' daily lives, which lasted 10 days \cite{de2024investigating, de2023daily}. However, this study is a secondary analysis that focuses solely on data collected during the real-world phase of the original study. Participants were recruited in Leuven, Belgium, and the majority (60\%) were college students enrolled at KU Leuven. The participants had an average age of 24.5 years (SD = 3), with 38 individuals (84\%) identifying as female. Race and ethnicity were not collected for the Healthy Study.  

\textbf{EMA:}
Participants were given a dedicated smartphone to respond to 10 EMAs daily, which were sent at random intervals between 15 and 90 minutes apart. EMAs included 10 mood-assessment questions \cite{myin2001emotional}. The mood question set covered nine NA components (worried, stressed, anxious, annoyed, down, restless, tense, under pressure, ashamed) and four PA components (relaxed, cheerful, confident, in control). Each component was prefaced with “At the moment, I feel…” and participants rated their responses on a scale from 1 (not at all) to 7 (very much). Similarly to PANAS, PA (range: 4-28) and NA (range: 9-63) were calculated by summing the respective individual responses.

\textbf{Sensor Data and Processing:}
The participants wore a 2-electrode ECG chest patch and a wrist-worn sensor that collected electrodermal activity (EDA) at 256 Hz, skin temperature (ST) at 1 Hz, and accelerometer (ACC) at 32 Hz. 

We extracted features from ST, ECG, EDA, and ACC. The data is segmented into 30-minute segments, and features are extracted from these segments. We utilized the Python package biosppy \cite{biosppy} to process the ECG data. Biosppy uses a bandpass filter with 3 Hz and 45 Hz frequencies, a sampling rate of 256, and the Hamilton segmentation algorithm to extract R peaks. We then validated the R peaks using the algorithm by Hovsepian et al. \cite{hovsepian2015cstress}; this algorithm uses the criterion beat difference based on the maximum expected difference for a beat and the minimal artifact difference. We then used the Python package hrvanalysis \cite{pichot2016hrvanalysis} to extract heart rate and heart rate variability features. We used the method proposed by Taylor et al. to process and extract the statistical and wavelet features from EDA \cite{taylor2015automatic}. For ST, we filtered out outliers using the IQR and then extracted statistical features during each segment. We smoothed the accelerometer signal using a fourth-order 10-Hz low-pass Butterworth filter and obtained a baseline. Then, we used the sensormotion Python package \cite{sensormotion} to extract activity features. A detailed list of features can be found in Table ~\ref{tab:S7}.

\subsubsection{EMA Class Label Distribution}
\begin{figure}[!ht]
\includegraphics[width=\linewidth]{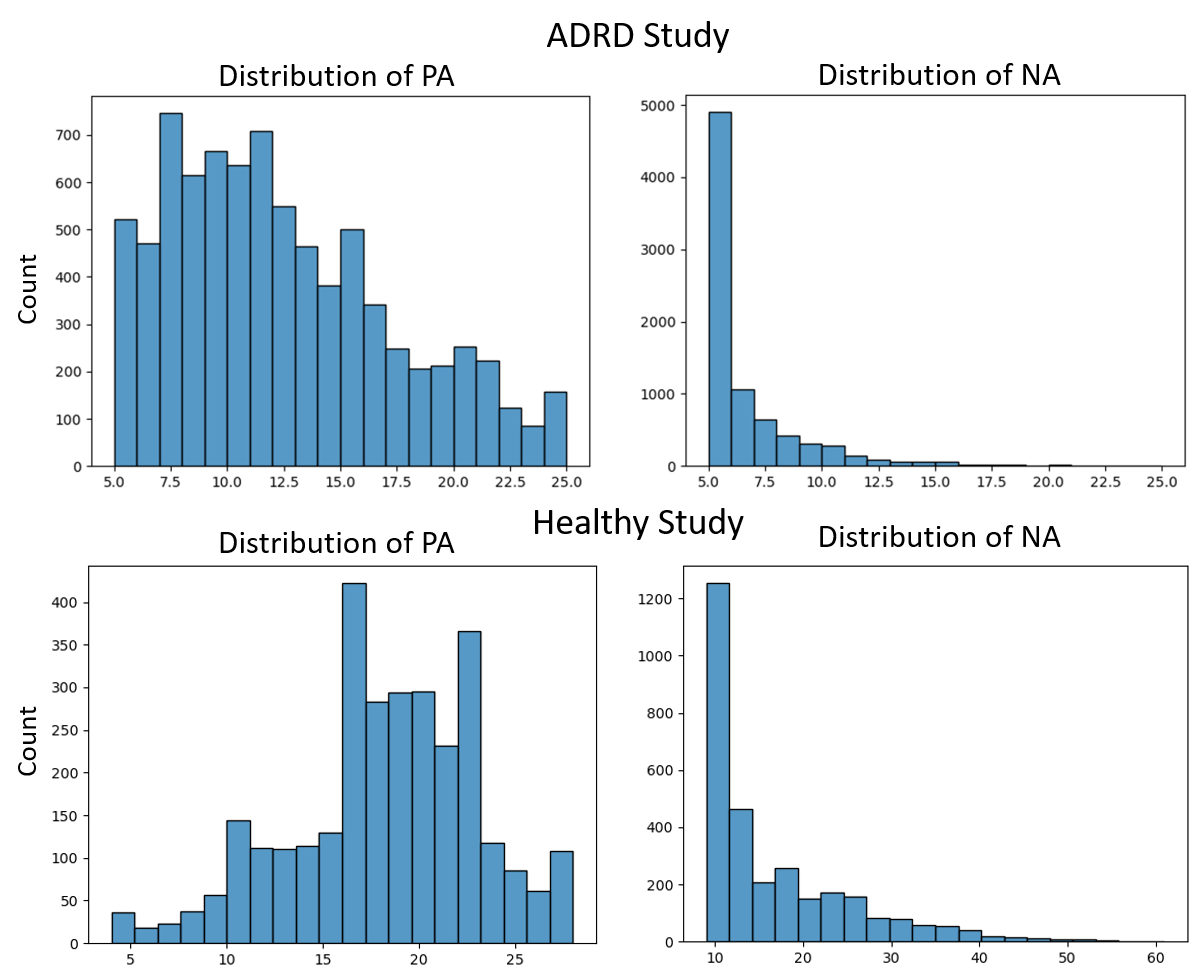}
\caption{Distribution of Positive Affect (PA) and Negative Affect (NA) among all participants.}
\label{fig:PAvsNA}
\end{figure} 

Figure \ref{fig:PAvsNA} illustrates the distribution of responses. PA shows a wide distribution range for both studies. In contrast, NA responses are strongly skewed with low variability, making accurate prediction challenging for an emotion model, as it lacks references to higher levels of NA. Given this distribution, our Emotion model will focus solely on PA scores.

\subsection{Evaluation}
\subsubsection{Receptivity and Emotion Model Settings and Evaluation}
We employ a cross-validation approach, which we refer to as semi-personalized. The semi-personalized approach begins similarly to Leave-One-Subject-Out (LOSO), but gradually incorporates personalized data. The first day is tested using a model trained only with other participants’ data. For each of the following days, the previous days from the test participant are added to the training set. This method replicates how a researcher may personalize an emotion or receptivity prediction model. This method outperforms a purely personalized approach due to the insufficient data available from each individual participant. We utilize a min-max normalization method. On the first day, we normalize the features using the data from the other participants. Subsequent days are normalized with respect to the participant's previous days of data. 

We test the performance of several machine learning algorithms and a random baseline model. For the receptivity model, we calculate the weighted F1 score (harmonic mean of precision and recall), Accuracy (proportion of correctly classified instances), and weighted Precision (weighted average of precision for both classes). For the emotion recognition model, we calculate Root Mean Square Error (RMSE) and $R^2$. Additionally, we conduct an ANOVA to compare performance metrics across the different machine learning algorithms, evaluated at the participant level. If the ANOVA test indicated a significant effect, we conducted post hoc Tukey tests to identify specific differences between the algorithms.

\subsubsection{Trigger Evaluation}

In an ideal situation, the multi-objective function would be tested against a control (randomly sending EMAs throughout the day) in a real-world study. However, to demonstrate the feasibility of our system before deploying it in a real-world setting, we test the multi-objective function offline using the high-resolution sensing and EMA data described in previous sections.

\subsubsection{Analyzing the output of the multi-objective function ($J$): } This evaluation method assesses the relationship between $J$ (Eq. \ref{opt}) and reported emotional state and receptivity. By examining these relationships, we can assess whether the proposed trigger would likely prompt users when they are more receptive or during emotional states that are more beneficial to model development.

\subsubsection{RQ1 (Would the proposed multi-objective function send EMAs when people are more responsive?):} 
To address \textit{RQ1}, we use a mixed effect model to determine the statistical relationship between $J$ and receptivity labels, where the response label is the dependent variable, $J$ is the fixed effect, and participant ID is the random effect. This will demonstrate how the output of the multi-objective function ($J$) behaves during responsive and non-responsive states, while accounting for individual differences across participants. We also utilize a repeated-measure ANOVA. However, since we are collecting repeated measures and repeated outcomes, the dependent variable ($J$) will be aggregated among responses and non-responses for each participant. Using these two statistical methods, we show that $J$ is higher during receptive time points, indicating the proposed objective function prioritizes moments when participants are more likely to respond.

\subsubsection{RQ2 (Would the proposed multi-objective function send EMAs when participants are experiencing emotional states that are less observed?):}
We employed a linear mixed-effect model to address \textit{RQ2}. Rather than directly analyzing the relationship between PA and $J$, we investigated the relationship between $J$ and the absolute standardized score of PA, representing how far each value is from the participant's mean. To focus on the magnitude of deviations, regardless of direction, we used the absolute value of the z-scores. Finally, we fit a mixed-effect model with $J$ as the fixed effect, participant ID as the random effect, and the absolute z-scores as the dependent variable. 

We also visualize the relationship between reported PA and $J$ by plotting the distribution of PA (y-axis) against a fitted line representing the average value of $J$ (x-axis) during varying levels of PA. This highlights the potential of the multi-objective function to prioritize EMA delivery during moments of higher model uncertainty. Emotion scores further from the mean are more likely to be underrepresented in the dataset, are generally harder to predict, and are more beneficial to incremental learning. 

\subsubsection{RQ3 (To what extent would the multi-objective function improve participant compliance and enhance the distribution of captured emotional responses compared to traditional EMA delivery methods?):}

To assess the potential impact of the multi-objective function on EMA delivery and data collection, we compare its performance with that of a traditional random sampling approach across multiple time windows. Each day for each participant is divided into five evenly spaced three-hour windows, and within each window, a delivery time point is selected using the two EMA delivery methods. The two EMA delivery strategies are: (1) the Smart Trigger, which selects the delivery time point that maximizes the output of the multi-objective function ($J$), and (2) a random trigger, which selects a delivery time point at random.

For the random trigger, the selected time point is chosen randomly during the scheduled window. In contrast, the Smart Trigger selected the time point that maximized the value of $J$ (Eq. \ref{opt}). We utilize our receptivity and emotion recognition models to obtain predicted values for responsiveness (response or non-response) and emotion (PA score) at each selected time point. We then calculate each participant’s response rate using both triggers. A paired t-test is used to assess whether the difference in participant response rates between the two EMA triggers is statistically significant.

Additionally, we analyze the distribution of predicted emotional states captured by each strategy. For both the Smart Trigger and the random trigger, we predict emotion scores at each selected time point to compare the distributions of participants' predicted emotions captured by each method. This comparison is done using the Kolmogorov–Smirnov (KS) Test, which assesses whether the distributions differ significantly. We also conducted a paired t-test to compare the within-participant variance in predicted emotions between the two triggers. This allows us to determine whether one strategy captures a broader range of emotions within individuals. 
\section{Results}

\subsection{Model Evaluation}
The results of our receptivity model for the ADRD and Healthy datasets using the semi-personalized cross-validation approach are shown in Table~\ref{tab:recpMetric}. We also analyzed the model performance using a Leave-One-Subject-Out (LOSO) cross-validation approach; for more information, see Table \ref{tab:S2}. The neural network is the best overall performer for the ADRD dataset, achieving higher accuracy and F1 scores compared to the other algorithms. However, this performance comes at the cost of reduced precision in identifying the positive (Responsive) class. We conducted an ANOVA to compare model performance across participants and found no significant differences in F1 score or precision. However, there was a significant effect for accuracy (F = 8.2, p $<$ 0.001).  A post hoc Tukey test revealed that the neural network's accuracy was significantly different from all other models, and that each model, excluding SVM, was statistically different from the baseline.

For the Healthy dataset, the random forest model performs best overall, achieving the highest accuracy and F1 scores. While its precision is comparable to the other models, it offers a clear advantage in overall predictive performance. In the Healthy dataset, the ANOVA results showed that only Accuracy demonstrated a significant difference across algorithms (F = 8.0, p $<$ 0.001). The post hoc Tukey test revealed that all algorithms except SVM were significantly different from the Baseline. Additionally, the neural network and random forest models differed significantly from all other algorithms, except for each other.


Table~\ref{tab:affMetric} shows RMSE results for the personalized models predicting emotional state. For results using the LOSO cross-validation approach, see Table \ref{tab:S4}. The neural network outperforms both linear regression and the baseline model, achieving an RMSE of approximately 3.5 for the ADRD dataset and 4.1 for the Healthy dataset. It is also the only model to achieve a positive $R^2$, with values of 0.48 for the ADRD study and 0.27 for the Healthy study. All other models produced negative $R^2$ values, indicating performance worse than predicting the mean. This was partly due to low variability in reported emotions for some participants. For both datasets, RMSE was statistically different across algorithms (ADRD Dataset; F = 13.9, p $<$ 0.001, Healthy Dataset; F = 11.5, p $<$ 0.001). For both datasets, the Tukey test showed that the neural network’s RMSE was significantly different from the Linear Regression and Baseline models.


\begin{table}
    \centering

    \begin{tabular}{|c|c|c|c|}
    \hline
     \multicolumn{4}{|c|}{ADRD Study} \\
    \hline
     Algorithm & F1 Score & Accuracy & Precision \\
     \hline
     Random Forest & 0.62(0.20) & 0.70(0.17) & 0.69(0.22) \\
     \hline
     Neural Network & 0.67(0.16) &  0.73(0.14)& 0.64(0.23)\\
     \hline
     SVM & 0.59 (0.07)&  0.58 (0.06)& 0.70 (0.24)\\
     \hline
     Naive Bayes & 0.61(0.17) & 0.64(0.15) & 0.69(0.22)\\
     \hline
     Baseline & 0.59(0.09) & 0.56(0.07) &0.68(0.22)\\
     \hline
     \hline

     \multicolumn{4}{|c|}{Healthy Study}\\
     \hline
     Algorithm & F1 Score & Accuracy & Precision \\
     \hline
     Random Forest  & 0.75 (0.09)& 0.64 (0.08)&0.78 (0.13)\\
     \hline
     Neural Network & 0.71(0.15) & 0.75(0.15) & 0.71(0.14) \\
     \hline
     SVM & 0.69 (0.27) & 0.78 (0.21) & 0.78 (0.21)\\
     \hline
     Naive Bayes &0.76 (0.11)& 0.67 (0.10)& 0.77 (0.14)\\
     \hline
     Baseline & 0.62 (0.07)& 0.73 (0.15) & 0.73 (0.15)\\
     \hline
     
\end{tabular}
\caption{Results of predicting receptivity using a Semi-Personalized cross-validation approach across multiple learning algorithms.}
\label{tab:recpMetric}
\end{table}


\begin{table}
    \centering

    \begin{tabular}{|c|c|c|c|c|}
    \hline
    Emotion Models& \multicolumn{2}{c|}{ADRD Study}& \multicolumn{2}{c|}{Healthy Study}\\
    \hline
     Algorithm & RMSE & $R^2$ & RMSE & $R^2$ \\
     \hline
     \hline
      Neural Network & 3.56 (0.2) & 0.48 (0.1) & 4.15 (0.02) & 0.27 (0.1) \\
     \hline
     Linear Regression
      & 4.6(2.0)& -4.7(11.0)&4.98(1.81)&-9.41 (6.32)\\
     \hline
     Baseline & 6.9(1.4)& -11.1(28.7)&6.73 (1.26)& -1.04 (0.89)\\
     \hline
     
\end{tabular}
\caption{Root Mean Square Error (RMSE) and $R^2$ metrics obtained from emotion prediction using the ADRD and Healthy dataset and a semi-personalized cross-validation approach.}
\label{tab:affMetric}
\end{table}

The results discussed in this section utilize data segmented into 30-minute-long windows. Model performance across multiple window lengths is detailed in \ref{tab:S3} for the receptivity model and in \ref{tab:S5} for the Emotion model. These findings indicate that 30-minute segments outperform other window lengths, especially the 10- and 15-minute windows. While performance differences between the 30- and 60-minute windows are smaller, longer segments place the data further from the actual EMA response time.

\subsection{Trigger Evaluation}

\subsubsection{Relationship between the output of the multi-objective function $J$ and Receptivity (\textit{RQ1})} 

\begin{figure}[!ht]
\includegraphics[width=0.85\linewidth]{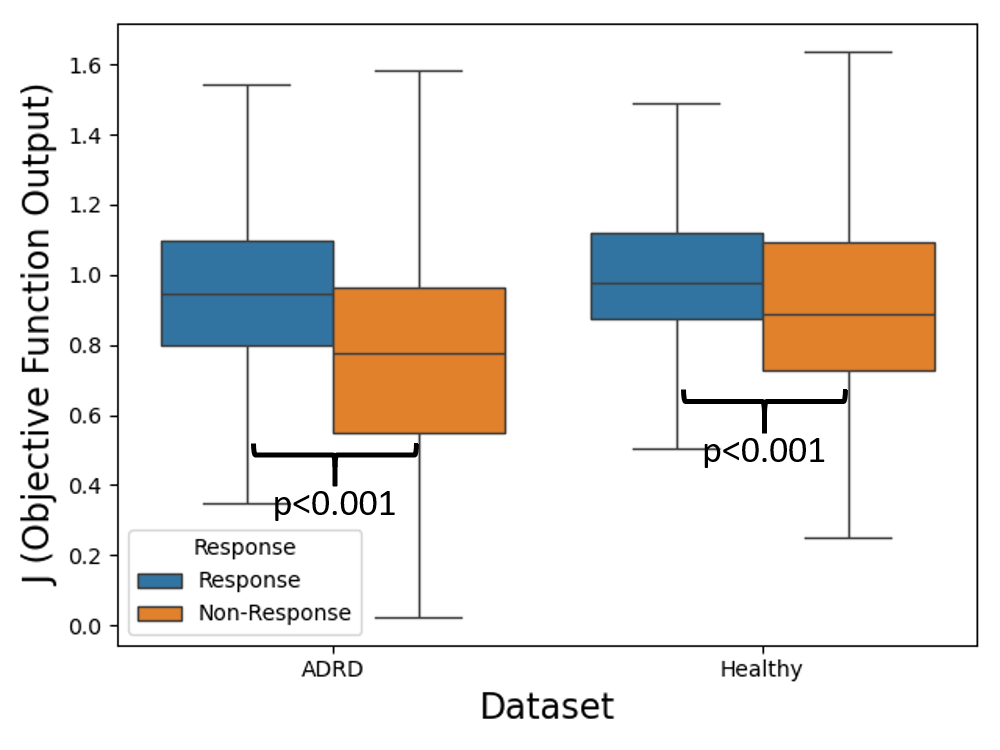}
\caption{Differences in $J$ (Equation \ref{opt}) during responses and non-responses for each population.}
\label{fig:ResponsevJ}
\end{figure}

Figure~\ref{fig:ResponsevJ} illustrates the differences in $J$ between responses and non-responses for each study. The observed differences suggest that an EMA trigger using this multi-objective function would likely send EMAs at times when participants are responsive. Using a mixed linear model, we found that $J$ was a significant predictor of receptivity for both the ADRD caregiver and the Healthy population. A few participants from each study were excluded from this analysis because their labeled data consisted entirely of responses or entirely of non-responses.

\textbf{ADRD Study:}  The estimated coefficient for $J$ was 0.04 (\emph{p} $<$ 0.001, 95\% CI: [0.027:0.047]).  The intercept was also significant (\emph{z} = 0.89, \emph{p} $<$ 0.001), highlighting a strong baseline likelihood of response across participants. The random effect variance for participants was 0.04, indicating some variability between participants in their response tendencies.

\textbf{Healthy Study:}  The estimated coefficient for $J$ was 0.03 (\emph{p} $< 0.001$, 95\% CI: [0.027:0.037]), suggesting a positive relationship between $J$ and the probability of response. The intercept was also significant (\emph{z} = 0.98, \emph{p} $<$ 0.001), highlighting a strong baseline likelihood of response across participants. The random effect variance for participants was 0.02, indicating some variability between participants in their response tendencies. 

We also evaluated the statistical difference in $J$ between responses and non-responses using repeated measures ANOVA. For both studies, we can visually notice the differences in $J$ during responses and non-responses. However, the repeated measure ANOVA takes into account the participant-specific differences. For the ADRD study, the results of the repeated-measures ANOVA indicate a statistically significant effect on responsiveness (\emph{F}=4.5, \emph{p} = 0.04), suggesting that $J$ differs significantly between instances of responses and non-responses. For the Healthy study, we found that this relationship was not significant. This is partly due to the number of participants in the Healthy dataset. Since we aggregate values of $J$ during responses and non-responses, we effectively reduce the number of observations to the number of participants, which may result in insignificant findings.

\subsubsection{Relationship between  the output of the multi-objective function $J$ and Reported Emotional State (\textit{RQ2})}

\begin{figure}[!ht]
     \centering
     \begin{subfigure}{0.49\textwidth}
         \centering
         \caption{ADRD Caregiver Study}
         \includegraphics[width=7.1cm, height=5.2cm]{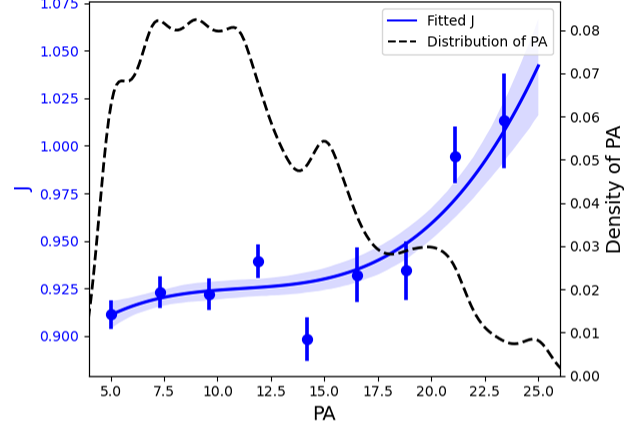}
         \label{fig:PAvJADRD}
     \end{subfigure}
     \begin{subfigure}{0.49\textwidth}
         \centering
         \caption{Healthy Study}
         \includegraphics[width=7.1cm, height=5.2cm]{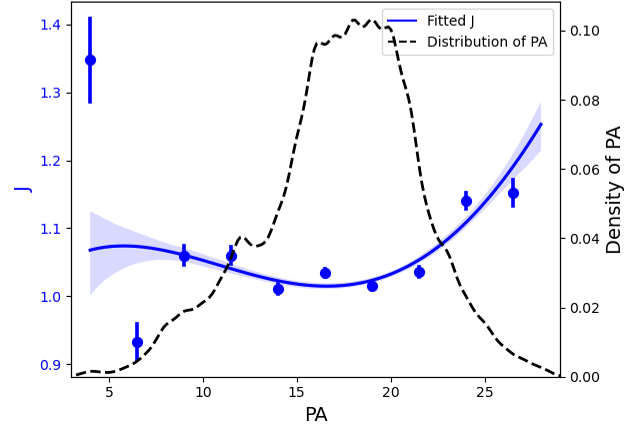}
         \label{fig:PAvJHealthy}
     \end{subfigure}
     \caption{The dashed line represents the distribution of collected positive affect (PA) labels during the day. Note that the y-axis does not correspond with the distribution of PS. The solid line represents the relationship between PA and $J$ ( Equation \ref{opt})}
     \label{fig:1}
\end{figure}

Figure~\ref{fig:PAvJADRD} illustrates the relationship between $J$ and reported PA for the ADRD caregiver study, and Figure~\ref{fig:PAvJHealthy} illustrates the same relationship for the Healthy study. Notably, $J$ is larger at PA scores where there are fewer observations, as seen on the edges of the distribution. This suggests that an EMA trigger using the multi-objective function would prioritize sending EMAs when participants are experiencing less frequently observed emotions, potentially capturing atypical emotional states that are more beneficial for model development.

We also found that the relationship we described between $J$ and the distribution of PA (Figure~\ref{fig:1}) for both studies is statistically meaningful. As mentioned, we converted the reported PA values to participant-specific absolute z-scores. This means that our mixed-effects model is examining the relationship between $J$ and the degree to which the value deviates from the participant's average reported PA score. For the \textbf{ADRD study}, the effect of J on the absolute z-score was statistically significant (\emph{p} $<$ 0.001), with a coefficient of 0.08 (95\% CI: [0.095, 0.11]). For the \textbf{Healthy Study}, the effect of J on absolute z-scores was statistically significant (\emph{p} $<$ 0.001), with a coefficient of 0.12 (95\% CI: [0.09, 0.14]). 

The results from both studies indicate that higher values of $J$ were associated with PA values further from the mean. And because of the distribution of reported PA, that would mean that values of $J$ are generally larger during reported emotions that are less represented in the dataset. The resulting coefficients for both studies are relatively small, which can be explained by the range and distribution of the absolute z-scores.

\subsubsection{Evaluating the Impact of the multi-objective function on Compliance and sampled Emotions Compared to Traditional EMA Delivery (\textit{RQ3})} 

The results of the simulated trigger indicate that receptivity rates were significantly higher when analyzing time points selected by the Smart Trigger compared to those chosen by the random trigger. Across the 73 ADRD participants with labeled data, the average receptivity rate for the random trigger was 0.82 (SD = 0.28), while the Smart Trigger achieved a higher rate of 0.93 (SD = 0.20). A paired t-test revealed that this difference in participant receptivity rates was statistically significant (t = 2.47, p = 0.01). For the Healthy dataset, the average receptivity rate was 0.87 (SD = 0.19) for the random trigger and 0.94 (SD = 0.10) for the Smart Trigger. These findings were also statistically significant based on a paired t-test (t = 2.14, p = 0.04).

The findings also suggest meaningful differences in emotional states at the time points selected by the two EMA triggers. For the ADRD population, a comparison of the PA score distributions revealed statistically significant differences, as determined by the KS test. However, the test statistic indicates that these differences were relatively subtle. On average, PA values at time points selected by the Smart Trigger were 13.7 (SD = 4.2), compared to 13.4 (SD = 4.0) for the random trigger. Notably, both averages are higher than the overall average PA score in the original dataset, which was 11.8 (SD = 4.9). For the Healthy dataset, a comparison of PA score distributions revealed statistically significant differences, as determined by the KS test. On average, PA values at time points selected by the Smart Trigger were 16.9 (SD = 4.1), while those selected by the random trigger averaged slightly higher at 17.1 (SD = 3.5). 

Finally, we observed a significant difference in participant-specific emotion score variance between the Smart Trigger and the random trigger for the ADRD population, as shown by a paired t-test (t = -1.97, p $<$ 0.05). On average, the within-participant variance was slightly higher for the Smart Trigger (M = 3.04, SD = 0.92) than for the random trigger (M = 2.83, SD = 0.93). For the Healthy population, we also found a significant difference in participant-specific emotion score variance between the Smart Trigger and the random trigger, as indicated by a paired t-test (t = -2.44, p $<$ 0.05). The average within-participant variance was higher for the Smart Trigger (M = 2.76, SD = 1.45) compared to the random trigger (M = 2.15, SD = 1.01). These results suggest that the Smart Trigger may capture a wider range of emotional states within individuals.

\section{Discussion}

\subsection{Model Performance} 

The performance of both the receptivity and emotion recognition models aligns with results reported in prior research. Notably, the receptivity model surpasses the Just-In-Time receptivity approach proposed by Kunzler et al., who developed a similar system for delivering interventions at moments when participants are most likely to respond \cite{kunzler2019exploring}. In their study, they achieved an F1-score of 0.31. Our receptivity model also outperforms several other works that aim to send messages, specifically interventions, at more responsive time points \cite{mishra2017investigating, mishra2021detecting}. However, none of the mentioned studies incorporate time-series wearable data; instead, they focus on passively collected contextual information from participants' smartphones. This underscores the value of time-series wearable data in accurately predicting a participant’s receptivity state. Directly comparing the performance of our emotion recognition model to previous work is challenging, primarily because our approach focuses on regression rather than binary or multi-class classification. Moreover, among the few studies that employ regression-based emotion recognition, most rely on data from facial expressions captured by cameras or participant speech or text, rather than wearable sensor data.

\subsection{Implications of the Multi-Objective Function}
This paper presents a new method that can significantly improve data collection in mHealth studies, particularly for those developing machine learning (ML) models. While our community has continuously strived to make the objective measures (wearable devices) more accurate, efficient, and robust, comparatively less attention has been given to advancing the subjective data collection methods, such as EMAs. Our evaluation demonstrates the potential of using machine learning techniques to distribute EMAs. Many mHealth studies, particularly those focused on well-being, rely on subjective measures that can only be sampled a few times a day. Maximizing the utility of each EMA can enhance the quality of our data and models, allowing us to intervene when needed. 

The objective function we have presented is modular, wherein any construct can be seamlessly integrated into Equation ~\ref{opt}. Its inherent flexibility allows for tailored adaptations to suit any population. The devices, sensors, and EMA constructs should be aligned with the population's characteristics. Specifically, for leveraging uncertainty, it is beneficial to collect data that is known to relate to the construct a researcher aims to predict. Additionally, the weights in Equation \ref{opt} must be adjusted based on the unique needs and differences of each population. For example, populations that struggle with EMA adherence might benefit from a higher weight of receptivity ($w_R$). These weights could also be adjusted adaptively using a reinforcement learning approach. In this method, rewards are defined for EMAs that prompt a response or yield valuable information for the model, while penalties are applied to non-responses. The model would then explore various weight combinations for uncertainty and receptivity and, over time, converge on the optimal weights for each specific participant or population.

\subsection{Limitations}

\subsubsection{Evaluation and Study-Specific Limitations}

The results demonstrated in this paper indicate that the trigger using the multi-objective function prefers sending EMAs when participants are more responsive and experiencing less common emotional states. However, these findings were validated offline and are therefore speculative. To better understand the proposed objective function, a study must be conducted with an experimental group receiving the EMA based on the proposed multi-objective function and a control group receiving the EMAs at random. Another major limitation of this paper is that the data obtained from Fitbit for the ADRD dataset may not be the optimal set of signals for momentary emotion prediction. Most researchers developing momentary emotion prediction and receptivity models use high-frequency signals, such as ECG or EDA, similar to the signals collected in the healthy study.  However, the Fitbit device is relatively easy for any population to use, regardless of age and technological intelligence.  Furthermore, our results may be influenced by population bias. By testing the trigger on two distinct populations, including a healthy general population, we aimed to improve the generalizability of our findings to other groups.

However, the differences in data collection methods between the two populations can also present a potential limitation. While these differences demonstrate the generalizability of the multi-objective function, they also introduce variability. Specifically, the data collected and the study duration differ: for ADRD caregivers, we collect minute-by-minute Fitbit data and contextual data from the AWARE app over three weeks, whereas for the healthy population, we gather high-frequency ST, EDA, ECG, and ACC data over 10 consecutive days. While these variations highlight that the multi-objective function can be applied across diverse data collection protocols, they may also pose challenges in interpreting results, as differences in data resolution and study duration can influence model performance and impact efficacy.

\subsubsection{Distribution of Class Labels}

The distribution of PA (our target variable) in both studies shows a central tendency, with occurrences being more frequent near the mean and decreasing as values deviate further from it. This allows us to assume that model uncertainty will be greater for emotions further from the mean, as model uncertainty is generally higher for less common occurrences. Additionally, participants reported varying intensities of positive emotions, unlike NA. A higher variance is needed to develop robust machine learning models that accurately predict emotions. When testing this trigger using NA as the target variable, we were unable to develop a model that was accurate enough to meet the project's needs. Nevertheless, the model would find that most negative emotions would have higher values of $J$, albeit there is only a small subset of higher values of reported NA. This lack of higher-intensity negative emotions may not solely reflect participants' true emotions but could instead stem from psychological factors influencing their willingness to report intense negative emotions. Social desirability bias may be one factor influencing this lack of variability in reported NA; Bergen et al. describes social desirability bias as the tendency to present oneself in a more socially acceptable manner \cite{bergen2020everything}.

While our findings demonstrate success in targeting moments of responsiveness and model uncertainty, they cannot fully address external psychological factors, such as social desirability bias, that influence participants' reported emotions. However, due to the modularity of the objective function, additional variables could be incorporated to account for such factors. For instance, introducing a third variable that measures the difference between the model's emotion prediction and the participant's average reported emotion could help identify moments when participants are more likely to deviate from their typical patterns of reporting. This approach could increase the likelihood of capturing genuine emotional states, including those participants who may otherwise hesitate to disclose.

\subsection{Future Work}

In future work, we plan to investigate the weights of the proposed objective function. While decisions can be made before the study's deployment, they depend on the intended outcome and population specific to each study. Consequently, it may prove beneficial to explore adaptive weights that fit the evolving needs of each participant at different time points throughout a study. 

Finally, we would like to conduct a study to test the efficacy of a trigger using the proposed multi-objective function against a control group that received the EMAs randomly throughout the day and a comparison group that would receive EMAs based only on the predicted likelihood of a response. This study will evaluate the Trigger’s impact on EMA receptivity, model performance, and its effectiveness compared to a traditional method and an alternative method that does not account for model uncertainty.

\section{Conclusion}
In this study, we proposed a multi-objective function to better distribute EMAs based on receptivity and emotion prediction uncertainty to improve the timing of an EMA notification. Additionally, we demonstrated the effectiveness of the objective function using a novel evaluation approach and data, including emotional state responses, wearable physiological data, and contextual information collected from 91 ADRD caregivers and 45 healthy participants. These contributions collectively advance mHealth research by improving the quality of our subjective data.

Our proposed objective function enhances incremental learning by prioritizing the sending of EMAs when the emotion prediction model is uncertain, without compromising receptivity. We demonstrated the relationship between the output of the proposed trigger model ($J$) and PA, which would indicate that the objective function would send EMAs during emotional states that are less represented in the dataset (Figure \ref{fig:PAvJADRD} and \ref{fig:PAvJHealthy}). Our evaluation method provides insight into how the trigger may behave in a real-world setting. Future studies should explore how ML-based EMA triggers impact users’ reported emotional states, model performance, and other important factors such as engagement and participant fatigue.

In conclusion, based on the outcomes presented in this work, we believe that the demonstrated feasibility and potential enhancements in data collection support incorporating the proposed multi-objective function into a clinical study. While our findings demonstrate the potential of this system, further testing is necessary to establish the statistical significance between our system and a control. 
\section{Acknowledgments}

This work was supported by the National Institutes of Health under grants R01AG062690, K25AG070306, and R01DA059925, and by the National Science Foundation  (1840167 and 2047296).

\bibliographystyle{ACM-Reference-Format}
\bibliography{sample-base}


\begin{thebibliography}{55}


\ifx \showCODEN    \undefined \def \showCODEN     #1{\unskip}     \fi
\ifx \showDOI      \undefined \def \showDOI       #1{#1}\fi
\ifx \showISBNx    \undefined \def \showISBNx     #1{\unskip}     \fi
\ifx \showISBNxiii \undefined \def \showISBNxiii  #1{\unskip}     \fi
\ifx \showISSN     \undefined \def \showISSN      #1{\unskip}     \fi
\ifx \showLCCN     \undefined \def \showLCCN      #1{\unskip}     \fi
\ifx \shownote     \undefined \def \shownote      #1{#1}          \fi
\ifx \showarticletitle \undefined \def \showarticletitle #1{#1}   \fi
\ifx \showURL      \undefined \def \showURL       {\relax}        \fi
\providecommand\bibfield[2]{#2}
\providecommand\bibinfo[2]{#2}
\providecommand\natexlab[1]{#1}
\providecommand\showeprint[2][]{arXiv:#2}

\bibitem[Akhand et~al\mbox{.}(2021)]%
        {akhand2021facial}
\bibfield{author}{\bibinfo{person}{MAH Akhand}, \bibinfo{person}{Shuvendu Roy}, \bibinfo{person}{Nazmul Siddique}, \bibinfo{person}{Md~Abdus~Samad Kamal}, {and} \bibinfo{person}{Tetsuya Shimamura}.} \bibinfo{year}{2021}\natexlab{}.
\newblock \showarticletitle{Facial emotion recognition using transfer learning in the deep CNN}.
\newblock \bibinfo{journal}{\emph{Electronics}} \bibinfo{volume}{10}, \bibinfo{number}{9} (\bibinfo{year}{2021}), \bibinfo{pages}{1036}.
\newblock


\bibitem[Association(2019)]%
        {alzheimer20192019}
\bibfield{author}{\bibinfo{person}{Alzheimer's Association}.} \bibinfo{year}{2019}\natexlab{}.
\newblock \showarticletitle{2019 Alzheimer's disease facts and figures}.
\newblock \bibinfo{journal}{\emph{Alzheimer's \& dementia}} \bibinfo{volume}{15}, \bibinfo{number}{3} (\bibinfo{year}{2019}), \bibinfo{pages}{321--387}.
\newblock


\bibitem[Bari et~al\mbox{.}(2020)]%
        {bari2020automated}
\bibfield{author}{\bibinfo{person}{Rummana Bari}, \bibinfo{person}{Md~Mahbubur Rahman}, \bibinfo{person}{Nazir Saleheen}, \bibinfo{person}{Megan~Battles Parsons}, \bibinfo{person}{Eugene~H Buder}, {and} \bibinfo{person}{Santosh Kumar}.} \bibinfo{year}{2020}\natexlab{}.
\newblock \showarticletitle{Automated detection of stressful conversations using wearable physiological and inertial sensors}.
\newblock \bibinfo{journal}{\emph{Proceedings of the ACM on interactive, mobile, wearable and ubiquitous technologies}} \bibinfo{volume}{4}, \bibinfo{number}{4} (\bibinfo{year}{2020}), \bibinfo{pages}{1--23}.
\newblock


\bibitem[Bergen and Labont{\'e}(2020)]%
        {bergen2020everything}
\bibfield{author}{\bibinfo{person}{Nicole Bergen} {and} \bibinfo{person}{Ronald Labont{\'e}}.} \bibinfo{year}{2020}\natexlab{}.
\newblock \showarticletitle{“Everything is perfect, and we have no problems”: detecting and limiting social desirability bias in qualitative research}.
\newblock \bibinfo{journal}{\emph{Qualitative health research}} \bibinfo{volume}{30}, \bibinfo{number}{5} (\bibinfo{year}{2020}), \bibinfo{pages}{783--792}.
\newblock


\bibitem[Bidargaddi et~al\mbox{.}(2018)]%
        {bidargaddi2018prompt}
\bibfield{author}{\bibinfo{person}{Niranjan Bidargaddi}, \bibinfo{person}{Daniel Almirall}, \bibinfo{person}{Susan Murphy}, \bibinfo{person}{Inbal Nahum-Shani}, \bibinfo{person}{Michael Kovalcik}, \bibinfo{person}{Timothy Pituch}, \bibinfo{person}{Haitham Maaieh}, {and} \bibinfo{person}{Victor Strecher}.} \bibinfo{year}{2018}\natexlab{}.
\newblock \showarticletitle{To prompt or not to prompt? A microrandomized trial of time-varying push notifications to increase proximal engagement with a mobile health app}.
\newblock \bibinfo{journal}{\emph{JMIR mHealth and uHealth}} \bibinfo{volume}{6}, \bibinfo{number}{11} (\bibinfo{year}{2018}), \bibinfo{pages}{e10123}.
\newblock


\bibitem[Carreiras et~al\mbox{.}(15  )]%
        {biosppy}
\bibfield{author}{\bibinfo{person}{Carlos Carreiras}, \bibinfo{person}{Ana~Priscila Alves}, \bibinfo{person}{Andr\'{e} Louren\c{c}o}, \bibinfo{person}{Filipe Canento}, \bibinfo{person}{Hugo Silva}, \bibinfo{person}{Ana Fred}, {et~al\mbox{.}}} \bibinfo{year}{2015--}\natexlab{}.
\newblock \bibinfo{title}{{BioSPPy}: Biosignal Processing in {Python}}.
\newblock
\newblock
\urldef\tempurl%
\url{https://github.com/PIA-Group/BioSPPy/}
\showURL{%
\tempurl}
\newblock
\shownote{[Online; accessed <today>]}.


\bibitem[Chen et~al\mbox{.}(2023)]%
        {chen2023you}
\bibfield{author}{\bibinfo{person}{Yu-Chun Chen}, \bibinfo{person}{Yu-Jen Lee}, \bibinfo{person}{Kuei-Chun Kao}, \bibinfo{person}{Jie Tsai}, \bibinfo{person}{En-Chi Liang}, \bibinfo{person}{Wei-Chen Chiu}, \bibinfo{person}{Faye Shih}, {and} \bibinfo{person}{Yung-Ju Chang}.} \bibinfo{year}{2023}\natexlab{}.
\newblock \showarticletitle{Are you killing time? Predicting smartphone users’ time-killing moments via fusion of smartphone sensor data and screenshots}. In \bibinfo{booktitle}{\emph{Proceedings of the 2023 CHI Conference on Human Factors in Computing Systems}}. \bibinfo{pages}{1--19}.
\newblock


\bibitem[De~Calheiros~Velozo et~al\mbox{.}(2024)]%
        {de2024investigating}
\bibfield{author}{\bibinfo{person}{Joana De~Calheiros~Velozo}, \bibinfo{person}{Thomas Vaessen}, \bibinfo{person}{Stephan Claes}, {and} \bibinfo{person}{Inez Myin-Germeys}.} \bibinfo{year}{2024}\natexlab{}.
\newblock \showarticletitle{Investigating adverse daily life effects following a psychosocial laboratory stress task, and the moderating role of Psychopathology}.
\newblock \bibinfo{journal}{\emph{StreSS}} \bibinfo{volume}{27}, \bibinfo{number}{1} (\bibinfo{year}{2024}), \bibinfo{pages}{2380403}.
\newblock


\bibitem[De~Calheiros~Velozo et~al\mbox{.}(2023)]%
        {de2023daily}
\bibfield{author}{\bibinfo{person}{Joana De~Calheiros~Velozo}, \bibinfo{person}{Thomas Vaessen}, \bibinfo{person}{Ginette Lafit}, \bibinfo{person}{Stephan Claes}, {and} \bibinfo{person}{Inez Myin-Germeys}.} \bibinfo{year}{2023}\natexlab{}.
\newblock \showarticletitle{Is daily-life stress reactivity a measure of stress recovery? An investigation of laboratory and daily-life stress}.
\newblock \bibinfo{journal}{\emph{Stress and Health}} \bibinfo{volume}{39}, \bibinfo{number}{3} (\bibinfo{year}{2023}), \bibinfo{pages}{638--650}.
\newblock


\bibitem[Eisele et~al\mbox{.}(2020)]%
        {eisele2020effects}
\bibfield{author}{\bibinfo{person}{Gudrun Eisele}, \bibinfo{person}{Hugo Vachon}, \bibinfo{person}{Ginette Lafit}, \bibinfo{person}{Peter Kuppens}, \bibinfo{person}{Marlies Houben}, \bibinfo{person}{Inez Myin-Germeys}, {and} \bibinfo{person}{Wolfgang Viechtbauer}.} \bibinfo{year}{2020}\natexlab{}.
\newblock \showarticletitle{The effects of sampling frequency and questionnaire length on perceived burden, compliance, and careless responding in experience sampling data in a student population}.
\newblock \bibinfo{journal}{\emph{Assessment}} (\bibinfo{year}{2020}), \bibinfo{pages}{1073191120957102}.
\newblock


\bibitem[Ferreira et~al\mbox{.}(2015)]%
        {ferreira2015aware}
\bibfield{author}{\bibinfo{person}{Denzil Ferreira}, \bibinfo{person}{Vassilis Kostakos}, {and} \bibinfo{person}{Anind~K Dey}.} \bibinfo{year}{2015}\natexlab{}.
\newblock \showarticletitle{AWARE: mobile context instrumentation framework}.
\newblock \bibinfo{journal}{\emph{Frontiers in ICT}}  \bibinfo{volume}{2} (\bibinfo{year}{2015}), \bibinfo{pages}{6}.
\newblock


\bibitem[Galv{\~a}o et~al\mbox{.}(2021)]%
        {galvao2021predicting}
\bibfield{author}{\bibinfo{person}{Filipe Galv{\~a}o}, \bibinfo{person}{Soraia~M Alarc{\~a}o}, {and} \bibinfo{person}{Manuel~J Fonseca}.} \bibinfo{year}{2021}\natexlab{}.
\newblock \showarticletitle{Predicting exact valence and arousal values from EEG}.
\newblock \bibinfo{journal}{\emph{Sensors}} \bibinfo{volume}{21}, \bibinfo{number}{10} (\bibinfo{year}{2021}), \bibinfo{pages}{3414}.
\newblock


\bibitem[He et~al\mbox{.}(2017)]%
        {he2017emotion}
\bibfield{author}{\bibinfo{person}{Cheng He}, \bibinfo{person}{Yun-jin Yao}, {and} \bibinfo{person}{Xue-song Ye}.} \bibinfo{year}{2017}\natexlab{}.
\newblock \showarticletitle{An emotion recognition system based on physiological signals obtained by wearable sensors}.
\newblock In \bibinfo{booktitle}{\emph{Wearable sensors and robots}}. \bibinfo{publisher}{Springer}, \bibinfo{pages}{15--25}.
\newblock


\bibitem[Ho and Intille(2005)]%
        {ho2005using}
\bibfield{author}{\bibinfo{person}{Joyce Ho} {and} \bibinfo{person}{Stephen~S Intille}.} \bibinfo{year}{2005}\natexlab{}.
\newblock \showarticletitle{Using context-aware computing to reduce the perceived burden of interruptions from mobile devices}. In \bibinfo{booktitle}{\emph{Proceedings of the SIGCHI conference on Human factors in computing systems}}. \bibinfo{pages}{909--918}.
\newblock


\bibitem[Ho(2018)]%
        {sensormotion}
\bibfield{author}{\bibinfo{person}{Simon Ho}.} \bibinfo{year}{2018}\natexlab{}.
\newblock \bibinfo{title}{Sensor Motion}.
\newblock
\newblock
\urldef\tempurl%
\url{https://github.com/sho-87/sensormotion}
\showURL{%
\tempurl}
\newblock
\shownote{[Online; accessed <today>]}.


\bibitem[Hovsepian et~al\mbox{.}(2015)]%
        {hovsepian2015cstress}
\bibfield{author}{\bibinfo{person}{Karen Hovsepian}, \bibinfo{person}{Mustafa Al'Absi}, \bibinfo{person}{Emre Ertin}, \bibinfo{person}{Thomas Kamarck}, \bibinfo{person}{Motohiro Nakajima}, {and} \bibinfo{person}{Santosh Kumar}.} \bibinfo{year}{2015}\natexlab{}.
\newblock \showarticletitle{cStress: towards a gold standard for continuous stress assessment in the mobile environment}. In \bibinfo{booktitle}{\emph{Proceedings of the 2015 ACM international joint conference on pervasive and ubiquitous computing}}. \bibinfo{pages}{493--504}.
\newblock


\bibitem[Hu et~al\mbox{.}(2018)]%
        {hu2018scai}
\bibfield{author}{\bibinfo{person}{Long Hu}, \bibinfo{person}{Jun Yang}, \bibinfo{person}{Min Chen}, \bibinfo{person}{Yongfeng Qian}, {and} \bibinfo{person}{Joel~JPC Rodrigues}.} \bibinfo{year}{2018}\natexlab{}.
\newblock \showarticletitle{SCAI-SVSC: Smart clothing for effective interaction with a sustainable vital sign collection}.
\newblock \bibinfo{journal}{\emph{Future Generation Computer Systems}}  \bibinfo{volume}{86} (\bibinfo{year}{2018}), \bibinfo{pages}{329--338}.
\newblock


\bibitem[Huynh et~al\mbox{.}(2021)]%
        {huynh2021stressnas}
\bibfield{author}{\bibinfo{person}{Lam Huynh}, \bibinfo{person}{Tri Nguyen}, \bibinfo{person}{Thu Nguyen}, \bibinfo{person}{Susanna Pirttikangas}, {and} \bibinfo{person}{Pekka Siirtola}.} \bibinfo{year}{2021}\natexlab{}.
\newblock \showarticletitle{Stressnas: Affect state and stress detection using neural architecture search}. In \bibinfo{booktitle}{\emph{Adjunct Proceedings of the 2021 ACM International Joint Conference on Pervasive and Ubiquitous Computing and Proceedings of the 2021 ACM International Symposium on Wearable Computers}}. \bibinfo{pages}{121--125}.
\newblock


\bibitem[Intille et~al\mbox{.}(2016)]%
        {intille2016muema}
\bibfield{author}{\bibinfo{person}{Stephen Intille}, \bibinfo{person}{Caitlin Haynes}, \bibinfo{person}{Dharam Maniar}, \bibinfo{person}{Aditya Ponnada}, {and} \bibinfo{person}{Justin Manjourides}.} \bibinfo{year}{2016}\natexlab{}.
\newblock \showarticletitle{$\mu$EMA: Microinteraction-based ecological momentary assessment (EMA) using a smartwatch}. In \bibinfo{booktitle}{\emph{Proceedings of the 2016 ACM International Joint Conference on Pervasive and Ubiquitous Computing}}. \bibinfo{pages}{1124--1128}.
\newblock


\bibitem[Jameel et~al\mbox{.}(2022)]%
        {jameel2022mhealth}
\bibfield{author}{\bibinfo{person}{Leila Jameel}, \bibinfo{person}{Lucia Valmaggia}, \bibinfo{person}{Georgina Barnes}, {and} \bibinfo{person}{Matteo Cella}.} \bibinfo{year}{2022}\natexlab{}.
\newblock \showarticletitle{mHealth technology to assess, monitor and treat daily functioning difficulties in people with severe mental illness: A systematic review}.
\newblock \bibinfo{journal}{\emph{Journal of psychiatric research}}  \bibinfo{volume}{145} (\bibinfo{year}{2022}), \bibinfo{pages}{35--49}.
\newblock


\bibitem[Jones et~al\mbox{.}(2019)]%
        {jones2019compliance}
\bibfield{author}{\bibinfo{person}{Andrew Jones}, \bibinfo{person}{Danielle Remmerswaal}, \bibinfo{person}{Ilse Verveer}, \bibinfo{person}{Eric Robinson}, \bibinfo{person}{Ingmar~HA Franken}, \bibinfo{person}{Cheng K~Fred Wen}, {and} \bibinfo{person}{Matt Field}.} \bibinfo{year}{2019}\natexlab{}.
\newblock \showarticletitle{Compliance with ecological momentary assessment protocols in substance users: a meta-analysis}.
\newblock \bibinfo{journal}{\emph{Addiction}} \bibinfo{volume}{114}, \bibinfo{number}{4} (\bibinfo{year}{2019}), \bibinfo{pages}{609--619}.
\newblock


\bibitem[King et~al\mbox{.}(2024)]%
        {king2024investigating}
\bibfield{author}{\bibinfo{person}{Zachary~D King}, \bibinfo{person}{Han Yu}, \bibinfo{person}{Thomas Vaessen}, \bibinfo{person}{Inez Myin-Germeys}, {and} \bibinfo{person}{Akane Sano}.} \bibinfo{year}{2024}\natexlab{}.
\newblock \showarticletitle{Investigating Receptivity and Affect Using Machine Learning: Ecological Momentary Assessment and Wearable Sensing Study}.
\newblock \bibinfo{journal}{\emph{JMIR mHealth and uHealth}} \bibinfo{volume}{12}, \bibinfo{number}{1} (\bibinfo{year}{2024}), \bibinfo{pages}{e46347}.
\newblock


\bibitem[K{\"u}nzler et~al\mbox{.}(2019)]%
        {kunzler2019exploring}
\bibfield{author}{\bibinfo{person}{Florian K{\"u}nzler}, \bibinfo{person}{Varun Mishra}, \bibinfo{person}{Jan-Niklas Kramer}, \bibinfo{person}{David Kotz}, \bibinfo{person}{Elgar Fleisch}, {and} \bibinfo{person}{Tobias Kowatsch}.} \bibinfo{year}{2019}\natexlab{}.
\newblock \showarticletitle{Exploring the state-of-receptivity for mhealth interventions}.
\newblock \bibinfo{journal}{\emph{Proceedings of the ACM on Interactive, Mobile, Wearable and Ubiquitous Technologies}} \bibinfo{volume}{3}, \bibinfo{number}{4} (\bibinfo{year}{2019}), \bibinfo{pages}{1--27}.
\newblock


\bibitem[Kuo et~al\mbox{.}(2018)]%
        {kuo2018cost}
\bibfield{author}{\bibinfo{person}{Weicheng Kuo}, \bibinfo{person}{Christian H{\"a}ne}, \bibinfo{person}{Esther Yuh}, \bibinfo{person}{Pratik Mukherjee}, {and} \bibinfo{person}{Jitendra Malik}.} \bibinfo{year}{2018}\natexlab{}.
\newblock \showarticletitle{Cost-sensitive active learning for intracranial hemorrhage detection}. In \bibinfo{booktitle}{\emph{Medical Image Computing and Computer Assisted Intervention--MICCAI 2018: 21st International Conference, Granada, Spain, September 16-20, 2018, Proceedings, Part III 11}}. Springer, \bibinfo{pages}{715--723}.
\newblock


\bibitem[Li et~al\mbox{.}(2024)]%
        {li2024ask}
\bibfield{author}{\bibinfo{person}{Jixin Li}, \bibinfo{person}{Aditya Ponnada}, \bibinfo{person}{Wei-Lin Wang}, \bibinfo{person}{Genevieve Dunton}, {and} \bibinfo{person}{Stephen Intille}.} \bibinfo{year}{2024}\natexlab{}.
\newblock \showarticletitle{Ask Less, Learn More: Adapting Ecological Momentary Assessment Survey Length by Modeling Question-Answer Information Gain}.
\newblock \bibinfo{journal}{\emph{Proceedings of the ACM on interactive, mobile, wearable and ubiquitous technologies}} \bibinfo{volume}{8}, \bibinfo{number}{4} (\bibinfo{year}{2024}), \bibinfo{pages}{1--32}.
\newblock


\bibitem[Luxton et~al\mbox{.}(2011)]%
        {luxton2011mhealth}
\bibfield{author}{\bibinfo{person}{David~D Luxton}, \bibinfo{person}{Russell~A McCann}, \bibinfo{person}{Nigel~E Bush}, \bibinfo{person}{Matthew~C Mishkind}, {and} \bibinfo{person}{Greg~M Reger}.} \bibinfo{year}{2011}\natexlab{}.
\newblock \showarticletitle{mHealth for mental health: Integrating smartphone technology in behavioral healthcare.}
\newblock \bibinfo{journal}{\emph{Professional Psychology: Research and Practice}} \bibinfo{volume}{42}, \bibinfo{number}{6} (\bibinfo{year}{2011}), \bibinfo{pages}{505}.
\newblock


\bibitem[Mackinnon et~al\mbox{.}(1999)]%
        {mackinnon1999short}
\bibfield{author}{\bibinfo{person}{Andrew Mackinnon}, \bibinfo{person}{Anthony~F Jorm}, \bibinfo{person}{Helen Christensen}, \bibinfo{person}{Ailsa~E Korten}, \bibinfo{person}{Patricia~A Jacomb}, {and} \bibinfo{person}{Bryan Rodgers}.} \bibinfo{year}{1999}\natexlab{}.
\newblock \showarticletitle{A short form of the Positive and Negative Affect Schedule: Evaluation of factorial validity and invariance across demographic variables in a community sample}.
\newblock \bibinfo{journal}{\emph{Personality and Individual differences}} \bibinfo{volume}{27}, \bibinfo{number}{3} (\bibinfo{year}{1999}), \bibinfo{pages}{405--416}.
\newblock


\bibitem[Malik and Camm(1993)]%
        {HRV_GUIDE}
\bibfield{author}{\bibinfo{person}{Marek Malik} {and} \bibinfo{person}{A~John Camm}.} \bibinfo{year}{1993}\natexlab{}.
\newblock \showarticletitle{Components of heart rate variability—what they really mean and what we really measure}.
\newblock \bibinfo{journal}{\emph{The American journal of cardiology}} \bibinfo{volume}{72}, \bibinfo{number}{11} (\bibinfo{year}{1993}), \bibinfo{pages}{821--822}.
\newblock


\bibitem[Mehrotra et~al\mbox{.}(2015)]%
        {mehrotra2015designing}
\bibfield{author}{\bibinfo{person}{Abhinav Mehrotra}, \bibinfo{person}{Mirco Musolesi}, \bibinfo{person}{Robert Hendley}, {and} \bibinfo{person}{Veljko Pejovic}.} \bibinfo{year}{2015}\natexlab{}.
\newblock \showarticletitle{Designing content-driven intelligent notification mechanisms for mobile applications}. In \bibinfo{booktitle}{\emph{Proceedings of the 2015 ACM International Joint Conference on Pervasive and Ubiquitous Computing}}. \bibinfo{pages}{813--824}.
\newblock


\bibitem[Mishra et~al\mbox{.}(2021)]%
        {mishra2021detecting}
\bibfield{author}{\bibinfo{person}{Varun Mishra}, \bibinfo{person}{Florian K{\"u}nzler}, \bibinfo{person}{Jan-Niklas Kramer}, \bibinfo{person}{Elgar Fleisch}, \bibinfo{person}{Tobias Kowatsch}, {and} \bibinfo{person}{David Kotz}.} \bibinfo{year}{2021}\natexlab{}.
\newblock \showarticletitle{Detecting Receptivity for mHealth Interventions in the Natural Environment}.
\newblock \bibinfo{journal}{\emph{Proceedings of the ACM on Interactive, Mobile, Wearable and Ubiquitous Technologies}} \bibinfo{volume}{5}, \bibinfo{number}{2} (\bibinfo{year}{2021}), \bibinfo{pages}{1--24}.
\newblock


\bibitem[Mishra et~al\mbox{.}(2017)]%
        {mishra2017investigating}
\bibfield{author}{\bibinfo{person}{Varun Mishra}, \bibinfo{person}{Byron Lowens}, \bibinfo{person}{Sarah Lord}, \bibinfo{person}{Kelly Caine}, {and} \bibinfo{person}{David Kotz}.} \bibinfo{year}{2017}\natexlab{}.
\newblock \showarticletitle{Investigating contextual cues as indicators for EMA delivery}. In \bibinfo{booktitle}{\emph{Proceedings of the 2017 ACM International Joint Conference on Pervasive and Ubiquitous Computing and Proceedings of the 2017 ACM International Symposium on Wearable Computers}}. \bibinfo{pages}{935--940}.
\newblock


\bibitem[Mitenkova et~al\mbox{.}(2019)]%
        {mitenkova2019valence}
\bibfield{author}{\bibinfo{person}{Anna Mitenkova}, \bibinfo{person}{Jean Kossaifi}, \bibinfo{person}{Yannis Panagakis}, {and} \bibinfo{person}{Maja Pantic}.} \bibinfo{year}{2019}\natexlab{}.
\newblock \showarticletitle{Valence and arousal estimation in-the-wild with tensor methods}. In \bibinfo{booktitle}{\emph{2019 14th IEEE International Conference on Automatic Face \& Gesture Recognition (FG 2019)}}. IEEE, \bibinfo{pages}{1--7}.
\newblock


\bibitem[Morrison et~al\mbox{.}(2017)]%
        {morrison2017effect}
\bibfield{author}{\bibinfo{person}{Leanne~G Morrison}, \bibinfo{person}{Charlie Hargood}, \bibinfo{person}{Veljko Pejovic}, \bibinfo{person}{Adam~WA Geraghty}, \bibinfo{person}{Scott Lloyd}, \bibinfo{person}{Natalie Goodman}, \bibinfo{person}{Danius~T Michaelides}, \bibinfo{person}{Anna Weston}, \bibinfo{person}{Mirco Musolesi}, \bibinfo{person}{Mark~J Weal}, {et~al\mbox{.}}} \bibinfo{year}{2017}\natexlab{}.
\newblock \showarticletitle{The effect of timing and frequency of push notifications on usage of a smartphone-based stress management intervention: an exploratory trial}.
\newblock \bibinfo{journal}{\emph{PloS one}} \bibinfo{volume}{12}, \bibinfo{number}{1} (\bibinfo{year}{2017}), \bibinfo{pages}{e0169162}.
\newblock


\bibitem[Murray et~al\mbox{.}(2023)]%
        {murray2023prompt}
\bibfield{author}{\bibinfo{person}{Aja~Louise Murray}, \bibinfo{person}{Ruth Brown}, \bibinfo{person}{Xinxin Zhu}, \bibinfo{person}{Lydia~Gabriela Speyer}, \bibinfo{person}{Yi Yang}, \bibinfo{person}{Zhouni Xiao}, \bibinfo{person}{Denis Ribeaud}, {and} \bibinfo{person}{Manuel Eisner}.} \bibinfo{year}{2023}\natexlab{}.
\newblock \showarticletitle{Prompt-level predictors of compliance in an ecological momentary assessment study of young adults' mental health}.
\newblock \bibinfo{journal}{\emph{Journal of Affective Disorders}}  \bibinfo{volume}{322} (\bibinfo{year}{2023}), \bibinfo{pages}{125--131}.
\newblock


\bibitem[Myin-Germeys et~al\mbox{.}(2001)]%
        {myin2001emotional}
\bibfield{author}{\bibinfo{person}{Inez Myin-Germeys}, \bibinfo{person}{Jim van Os}, \bibinfo{person}{Joseph~E Schwartz}, \bibinfo{person}{Arthur~A Stone}, {and} \bibinfo{person}{Philippe~A Delespaul}.} \bibinfo{year}{2001}\natexlab{}.
\newblock \showarticletitle{Emotional reactivity to daily life stress in psychosis}.
\newblock \bibinfo{journal}{\emph{Archives of general psychiatry}} \bibinfo{volume}{58}, \bibinfo{number}{12} (\bibinfo{year}{2001}), \bibinfo{pages}{1137--1144}.
\newblock


\bibitem[Nagesh et~al\mbox{.}(2021)]%
        {nagesh2021transformers}
\bibfield{author}{\bibinfo{person}{Supriya Nagesh}, \bibinfo{person}{Alexander Moreno}, \bibinfo{person}{Stephanie~M Carpenter}, \bibinfo{person}{Jamie Yap}, \bibinfo{person}{Soujanya Chatterjee}, \bibinfo{person}{Steven~Lloyd Lizotte}, \bibinfo{person}{Neng Wan}, \bibinfo{person}{Santosh Kumar}, \bibinfo{person}{Cho Lam}, \bibinfo{person}{David~W Wetter}, {et~al\mbox{.}}} \bibinfo{year}{2021}\natexlab{}.
\newblock \showarticletitle{Transformers for prompt-level EMA non-response prediction}.
\newblock \bibinfo{journal}{\emph{arXiv preprint arXiv:2111.01193}} (\bibinfo{year}{2021}).
\newblock


\bibitem[Nalepa et~al\mbox{.}(2019)]%
        {nalepa2019analysis}
\bibfield{author}{\bibinfo{person}{Grzegorz~J Nalepa}, \bibinfo{person}{Krzysztof Kutt}, \bibinfo{person}{Barbara Gi{\.z}ycka}, \bibinfo{person}{Pawe{\l} Jemio{\l}o}, {and} \bibinfo{person}{Szymon Bobek}.} \bibinfo{year}{2019}\natexlab{}.
\newblock \showarticletitle{Analysis and use of the emotional context with wearable devices for games and intelligent assistants}.
\newblock \bibinfo{journal}{\emph{Sensors}} \bibinfo{volume}{19}, \bibinfo{number}{11} (\bibinfo{year}{2019}), \bibinfo{pages}{2509}.
\newblock


\bibitem[Ottenstein and Werner(2022)]%
        {ottenstein2022compliance}
\bibfield{author}{\bibinfo{person}{Charlotte Ottenstein} {and} \bibinfo{person}{Linda Werner}.} \bibinfo{year}{2022}\natexlab{}.
\newblock \showarticletitle{Compliance in ambulatory assessment studies: Investigating study and sample characteristics as predictors}.
\newblock \bibinfo{journal}{\emph{Assessment}} \bibinfo{volume}{29}, \bibinfo{number}{8} (\bibinfo{year}{2022}), \bibinfo{pages}{1765--1776}.
\newblock


\bibitem[Pejovic and Musolesi(2014)]%
        {pejovic2014interruptme}
\bibfield{author}{\bibinfo{person}{Veljko Pejovic} {and} \bibinfo{person}{Mirco Musolesi}.} \bibinfo{year}{2014}\natexlab{}.
\newblock \showarticletitle{InterruptMe: designing intelligent prompting mechanisms for pervasive applications}. In \bibinfo{booktitle}{\emph{Proceedings of the 2014 ACM International Joint Conference on Pervasive and Ubiquitous Computing}}. \bibinfo{pages}{897--908}.
\newblock


\bibitem[Phillips et~al\mbox{.}(2017)]%
        {phillips2017irregular}
\bibfield{author}{\bibinfo{person}{Andrew~JK Phillips}, \bibinfo{person}{William~M Clerx}, \bibinfo{person}{Conor~S O’Brien}, \bibinfo{person}{Akane Sano}, \bibinfo{person}{Laura~K Barger}, \bibinfo{person}{Rosalind~W Picard}, \bibinfo{person}{Steven~W Lockley}, \bibinfo{person}{Elizabeth~B Klerman}, {and} \bibinfo{person}{Charles~A Czeisler}.} \bibinfo{year}{2017}\natexlab{}.
\newblock \showarticletitle{Irregular sleep/wake patterns are associated with poorer academic performance and delayed circadian and sleep/wake timing}.
\newblock \bibinfo{journal}{\emph{Scientific reports}} \bibinfo{volume}{7}, \bibinfo{number}{1} (\bibinfo{year}{2017}), \bibinfo{pages}{1--13}.
\newblock


\bibitem[Pichot et~al\mbox{.}(2016)]%
        {pichot2016hrvanalysis}
\bibfield{author}{\bibinfo{person}{Vincent Pichot}, \bibinfo{person}{Fr{\'e}d{\'e}ric Roche}, \bibinfo{person}{S{\'e}bastien Celle}, \bibinfo{person}{Jean-Claude Barth{\'e}l{\'e}my}, {and} \bibinfo{person}{Florian Chouchou}.} \bibinfo{year}{2016}\natexlab{}.
\newblock \showarticletitle{HRVanalysis: a free software for analyzing cardiac autonomic activity}.
\newblock \bibinfo{journal}{\emph{Frontiers in physiology}}  \bibinfo{volume}{7} (\bibinfo{year}{2016}), \bibinfo{pages}{557}.
\newblock


\bibitem[Pielot et~al\mbox{.}(2017)]%
        {pielot2017beyond}
\bibfield{author}{\bibinfo{person}{Martin Pielot}, \bibinfo{person}{Bruno Cardoso}, \bibinfo{person}{Kleomenis Katevas}, \bibinfo{person}{Joan Serr{\`a}}, \bibinfo{person}{Aleksandar Matic}, {and} \bibinfo{person}{Nuria Oliver}.} \bibinfo{year}{2017}\natexlab{}.
\newblock \showarticletitle{Beyond interruptibility: Predicting opportune moments to engage mobile phone users}.
\newblock \bibinfo{journal}{\emph{Proceedings of the ACM on Interactive, Mobile, Wearable and Ubiquitous Technologies}} \bibinfo{volume}{1}, \bibinfo{number}{3} (\bibinfo{year}{2017}), \bibinfo{pages}{1--25}.
\newblock


\bibitem[Quintana et~al\mbox{.}(2012)]%
        {quintana2012heart}
\bibfield{author}{\bibinfo{person}{Daniel~S Quintana}, \bibinfo{person}{Adam~J Guastella}, \bibinfo{person}{Tim Outhred}, \bibinfo{person}{Ian~B Hickie}, {and} \bibinfo{person}{Andrew~H Kemp}.} \bibinfo{year}{2012}\natexlab{}.
\newblock \showarticletitle{Heart rate variability is associated with emotion recognition: Direct evidence for a relationship between the autonomic nervous system and social cognition}.
\newblock \bibinfo{journal}{\emph{International journal of psychophysiology}} \bibinfo{volume}{86}, \bibinfo{number}{2} (\bibinfo{year}{2012}), \bibinfo{pages}{168--172}.
\newblock


\bibitem[Rashid et~al\mbox{.}(2020)]%
        {rashid2020predicting}
\bibfield{author}{\bibinfo{person}{Haroon Rashid}, \bibinfo{person}{Sanjana Mendu}, \bibinfo{person}{Katharine~E Daniel}, \bibinfo{person}{Miranda~L Beltzer}, \bibinfo{person}{Bethany~A Teachman}, \bibinfo{person}{Mehdi Boukhechba}, {and} \bibinfo{person}{Laura~E Barnes}.} \bibinfo{year}{2020}\natexlab{}.
\newblock \showarticletitle{Predicting subjective measures of social anxiety from sparsely collected mobile sensor data}.
\newblock \bibinfo{journal}{\emph{Proceedings of the ACM on Interactive, Mobile, Wearable and Ubiquitous Technologies}} \bibinfo{volume}{4}, \bibinfo{number}{3} (\bibinfo{year}{2020}), \bibinfo{pages}{1--24}.
\newblock


\bibitem[Rintala et~al\mbox{.}(2020)]%
        {rintala2020momentary}
\bibfield{author}{\bibinfo{person}{Aki Rintala}, \bibinfo{person}{Martien Wampers}, \bibinfo{person}{Inez Myin-Germeys}, {and} \bibinfo{person}{Wolfgang Viechtbauer}.} \bibinfo{year}{2020}\natexlab{}.
\newblock \showarticletitle{Momentary predictors of compliance in studies using the experience sampling method}.
\newblock \bibinfo{journal}{\emph{Psychiatry research}}  \bibinfo{volume}{286} (\bibinfo{year}{2020}), \bibinfo{pages}{112896}.
\newblock


\bibitem[Rowland et~al\mbox{.}(2020)]%
        {rowland2020clinical}
\bibfield{author}{\bibinfo{person}{Simon~P Rowland}, \bibinfo{person}{J~Edward Fitzgerald}, \bibinfo{person}{Thomas Holme}, \bibinfo{person}{John Powell}, {and} \bibinfo{person}{Alison McGregor}.} \bibinfo{year}{2020}\natexlab{}.
\newblock \showarticletitle{What is the clinical value of mHealth for patients?}
\newblock \bibinfo{journal}{\emph{NPJ digital medicine}} \bibinfo{volume}{3}, \bibinfo{number}{1} (\bibinfo{year}{2020}), \bibinfo{pages}{4}.
\newblock


\bibitem[Sano and Picard(2013)]%
        {sano2013stress}
\bibfield{author}{\bibinfo{person}{Akane Sano} {and} \bibinfo{person}{Rosalind~W Picard}.} \bibinfo{year}{2013}\natexlab{}.
\newblock \showarticletitle{Stress recognition using wearable sensors and mobile phones}. In \bibinfo{booktitle}{\emph{2013 Humaine association conference on affective computing and intelligent interaction}}. IEEE, \bibinfo{pages}{671--676}.
\newblock


\bibitem[Sarker et~al\mbox{.}(2014)]%
        {sarker2014assessing}
\bibfield{author}{\bibinfo{person}{Hillol Sarker}, \bibinfo{person}{Moushumi Sharmin}, \bibinfo{person}{Amin~Ahsan Ali}, \bibinfo{person}{Md~Mahbubur Rahman}, \bibinfo{person}{Rummana Bari}, \bibinfo{person}{Syed~Monowar Hossain}, {and} \bibinfo{person}{Santosh Kumar}.} \bibinfo{year}{2014}\natexlab{}.
\newblock \showarticletitle{Assessing the availability of users to engage in just-in-time intervention in the natural environment}. In \bibinfo{booktitle}{\emph{Proceedings of the 2014 ACM international joint conference on pervasive and ubiquitous computing}}. \bibinfo{pages}{909--920}.
\newblock


\bibitem[Schwartz and Stone(1998)]%
        {schwartz1998strategies}
\bibfield{author}{\bibinfo{person}{Joseph~E Schwartz} {and} \bibinfo{person}{Arthur~A Stone}.} \bibinfo{year}{1998}\natexlab{}.
\newblock \showarticletitle{Strategies for analyzing ecological momentary assessment data.}
\newblock \bibinfo{journal}{\emph{Health Psychology}} \bibinfo{volume}{17}, \bibinfo{number}{1} (\bibinfo{year}{1998}), \bibinfo{pages}{6}.
\newblock


\bibitem[Taylor et~al\mbox{.}(2015)]%
        {taylor2015automatic}
\bibfield{author}{\bibinfo{person}{Sara Taylor}, \bibinfo{person}{Natasha Jaques}, \bibinfo{person}{Weixuan Chen}, \bibinfo{person}{Szymon Fedor}, \bibinfo{person}{Akane Sano}, {and} \bibinfo{person}{Rosalind Picard}.} \bibinfo{year}{2015}\natexlab{}.
\newblock \showarticletitle{Automatic identification of artifacts in electrodermal activity data}. In \bibinfo{booktitle}{\emph{2015 37th Annual International Conference of the IEEE Engineering in Medicine and Biology Society (EMBC)}}. IEEE, \bibinfo{pages}{1934--1937}.
\newblock


\bibitem[Vellinga et~al\mbox{.}(2020)]%
        {vellinga2020patients}
\bibfield{author}{\bibinfo{person}{Akke Vellinga}, \bibinfo{person}{Colum Devine}, \bibinfo{person}{Min~Yun Ho}, \bibinfo{person}{Colin Clarke}, \bibinfo{person}{Patrick Leahy}, \bibinfo{person}{Jane Bourke}, \bibinfo{person}{Declan Devane}, \bibinfo{person}{Sinead Duane}, {and} \bibinfo{person}{Patricia Kearney}.} \bibinfo{year}{2020}\natexlab{}.
\newblock \showarticletitle{What do patients value as incentives for participation in clinical trials? A pilot discrete choice experiment}.
\newblock \bibinfo{journal}{\emph{Research Ethics}} \bibinfo{volume}{16}, \bibinfo{number}{1-2} (\bibinfo{year}{2020}), \bibinfo{pages}{1--12}.
\newblock


\bibitem[Wen et~al\mbox{.}(2017)]%
        {wen2017compliance}
\bibfield{author}{\bibinfo{person}{Cheng K~Fred Wen}, \bibinfo{person}{Stefan Schneider}, \bibinfo{person}{Arthur~A Stone}, {and} \bibinfo{person}{Donna Spruijt-Metz}.} \bibinfo{year}{2017}\natexlab{}.
\newblock \showarticletitle{Compliance with mobile ecological momentary assessment protocols in children and adolescents: a systematic review and meta-analysis}.
\newblock \bibinfo{journal}{\emph{Journal of medical Internet research}} \bibinfo{volume}{19}, \bibinfo{number}{4} (\bibinfo{year}{2017}), \bibinfo{pages}{e132}.
\newblock


\bibitem[Wrzus and Neubauer(2023)]%
        {wrzus2023ecological}
\bibfield{author}{\bibinfo{person}{Cornelia Wrzus} {and} \bibinfo{person}{Andreas~B Neubauer}.} \bibinfo{year}{2023}\natexlab{}.
\newblock \showarticletitle{Ecological momentary assessment: A meta-analysis on designs, samples, and compliance across research fields}.
\newblock \bibinfo{journal}{\emph{Assessment}} \bibinfo{volume}{30}, \bibinfo{number}{3} (\bibinfo{year}{2023}), \bibinfo{pages}{825--846}.
\newblock


\bibitem[Yu and Sano(2023)]%
        {yu2023semi}
\bibfield{author}{\bibinfo{person}{Han Yu} {and} \bibinfo{person}{Akane Sano}.} \bibinfo{year}{2023}\natexlab{}.
\newblock \showarticletitle{Semi-Supervised Learning for Wearable-based Momentary Stress Detection in the Wild}.
\newblock \bibinfo{journal}{\emph{Proceedings of the ACM on Interactive, Mobile, Wearable and Ubiquitous Technologies}} \bibinfo{volume}{7}, \bibinfo{number}{2} (\bibinfo{year}{2023}), \bibinfo{pages}{1--23}.
\newblock


\bibitem[Zhao et~al\mbox{.}(2018)]%
        {zhao2018emotionsense}
\bibfield{author}{\bibinfo{person}{Bobo Zhao}, \bibinfo{person}{Zhu Wang}, \bibinfo{person}{Zhiwen Yu}, {and} \bibinfo{person}{Bin Guo}.} \bibinfo{year}{2018}\natexlab{}.
\newblock \showarticletitle{EmotionSense: Emotion recognition based on wearable wristband}. In \bibinfo{booktitle}{\emph{2018 IEEE SmartWorld, Ubiquitous Intelligence \& Computing, Advanced \& Trusted Computing, Scalable Computing \& Communications, Cloud \& Big Data Computing, Internet of People and Smart City Innovation (SmartWorld/SCALCOM/UIC/ATC/CBDCom/IOP/SCI)}}. IEEE, \bibinfo{pages}{346--355}.
\newblock


\end{thebibliography}

\appendix

\renewcommand{\thetable}{S\arabic{table}}

\section{ADRD Study Feature Extraction} \label{Dataset_App}
\textbf{Fitbit: }Four features are extracted from the Fitbit intraday data: heart rate, steps, sleep, and heart rate variability. To ensure the quality and reliability of the Fitbit data, we filtered out data points that fell outside our pre-defined expected ranges for steps (0-400), heart rate (40-200 beats per minute), and sleep stages. We also filtered out minutes where no data had been collected. This step was crucial because the Fitbit API may return a step count of 0, even when the device is charging or not worn. These minutes are identified by a lack of movement (0 steps) and the absence of heart rate (HR) or sleep data.

After validation, we segment the steps and heart rate data into 5- and 30-minute windows to extract statistical features such as mean, minimum, standard deviation, and more. Using segments shorter than 5 minutes poses challenges due to the low sampling rate available for Fitbit (minute-by-minute).
HRV features are derived from the R-R interval \cite{HRV_GUIDE} and calculated by Fitbit. The HRV features are only obtained when participants are asleep and are relatively still while sleeping. RMSSD (root mean square of successive differences between normal heartbeats) is one such feature, reflecting parasympathetic nervous system activity. Fitbit also calculated HF (0.15 to 0.40 Hz) and LF (0.04 to 0.15 Hz), which are frequency domain features. These HRV features can be predictive of stress and emotion \cite{quintana2012heart}.
When analyzing sleep data from Fitbit, we focus on several key features to assess the quality, consistency, and duration of the prior night's sleep. Fitbit provides intraday data categorized by the sleep state at each minute, allowing us to calculate various metrics. These metrics include sleep duration, time spent in different sleep stages, instances of waking up during the night, sleep efficiency, and sleep regularity. The sleep regularity index (SRI) quantifies the consistency of a participant's sleep pattern by assessing the likelihood of sleep occurring 24 hours apart \cite{phillips2017irregular}. SRI is scored between 0 and 100, where 0 represents random sleep patterns, while 100 indicates perfectly regular sleep (i.e., falling asleep and waking up at the same time each day). Sleep efficiency is another important metric calculated based on total sleep time, time spent in bed, and time awake before final awakening. A participant achieves 100\% sleep efficiency if they fall asleep immediately after going to bed and do not wake up during the night.
\textbf{Phone Data:} The Aware app collects four key data elements: location, call logs, message logs, and screen usage \cite{ferreira2015aware}. From these elements, we calculate seven features. Concerning participant privacy, we only gather metadata (e.g., timestamps of calls and messages, not phone or text conversations). 
We segment the data using the same segments from Fitbit (i.e., minute-by-minute and historical data). From there, we calculate many features, including messages received and sent, the length of conversations (phone and text), screen time, etc. When processing and extracting features from location, we cluster the locations and label the most frequent location as their home. We employ the k-means clustering technique, wherein the value of k is chosen for each participant through optimization of the silhouette score. However, we introduce a constraint to this selection process, stipulating that the number of clusters must exceed five. This constraint is implemented to ensure that our analysis maintains a level of granularity that prevents overgeneralization. We also clean the data by removing outliers or noisy locations that are unlikely or unavailable. 

\begin{table}
    \centering
    \setlength{\tabcolsep}{0.3\tabcolsep}
    
    \begin{tabular}{l|l}
    
          Demographic Characteristic & \hspace{0.3cm} ADRD Study (N=87) \\
          \hline
          \textbf{Age (years)} &   \\
          \hspace{0.5cm} Mean (SD) &  \hspace{1cm} 62 (11.8) \\
          \textbf{Sex, n (\%)} &   \\
          \hspace{0.5cm} Female & \hspace{1cm} 76 (87\%)\\
          \hspace{0.5cm} Male & \hspace{1cm} 11 (13\%)\\
          \textbf{Race, n (\%)} & \\
          \hspace{0.5cm} White & \hspace{1cm} 78 (90\%)\\
          \hspace{0.5cm} African American & \hspace{1cm} 8 (9\%) \\
          \hspace{0.5cm} American Indian, \\
          \hspace{0.75cm} Alaska Native & \hspace{1cm} 1 (1\%)\\
          \hspace{0.5cm} Native Hawaiian, & \\
          \hspace{0.75cm} Pacific Islander & \hspace{1cm} 1 (1\%)\\
          \hspace{0.5cm} Other & \hspace{1cm} 1 (1\%)\\
          \textbf{Ethnicity, n (\%)} & \\
          \hspace{0.5cm} Hispanic & \hspace{1cm} 3 (3\%)\\
          \hspace{0.5cm} Non-Hispanic & \hspace{1cm} 79 (91\%) \\
          \hspace{0.5cm} NA & \hspace{1cm}  6 (7\%)\\
    \end{tabular}
    \caption{Demographic information in ADRD Study Dataset}
    \label{tab:S1}
\end{table}


\begin{table}
    \centering

    \begin{tabular}{|c|c|c|c|}
    \hline
     \multicolumn{4}{|c|}{ADRD Study} \\
    \hline
     Algorithm & F1 Score & Accuracy & Precision \\
     \hline
     Random Forest & 0.59(0.23) & 0.67(0.21) & 0.69(0.22) \\
     \hline
     Neural Network & 0.59(0.24) & 0.68(0.22) & 0.67(0.24)\\
     \hline
     SVM & 0.56(0.06) & 0.51(0.06) & 0.71(0.23)\\
     \hline
     Naive Bayes & 0.60(0.18) & 0.63(0.16)&0.68(0.22)\\
     \hline
     Baseline & 0.56(0.09)& 0.55(0.08)& 0.66(0.21)\\
     \hline
     \hline

     \multicolumn{4}{|c|}{Healthy Study}\\
     \hline
     Algorithm & F1 Score & Accuracy & Precision \\
     \hline
     Random Forest  & 0.73(0.10) & 0.63(0.09) & 0.78(0.14)\\
     \hline
     Neural Network & 0.70(0.18)& 0.77(0.14)&0.72(0.15) \\
     \hline
     SVM & 0.45 (0.27)& 0.49 (0.16)& 0.62 (0.21)\\
     \hline
     Naive Bayes & 0.70 (0.11)& 0.69 (0.10)& 0.80 (0.13)\\
     \hline
     Baseline & 0.62 (0.11)& 0.62 (0.07)& 0.72 (0.14)\\
     \hline
     
\end{tabular}
\caption{Receptivity Model Performance Using LOSO Cross-Validation.}
\label{tab:S2}
\end{table}


\begin{table}
    \centering

    \begin{tabular}{|c|c|c||c|c|}
    \hline
     \multicolumn{4}{|c|}{ADRD Study}\\
    \hline
     Window Length  & Precision & Accuracy & F1-Score  \\
     \hline
     10-Minutes  &  0.51(0.09) & 0.62(0.17)  & 0.61(0.14)  \\
     \hline
     15-Minutes  &  0.48(0.09) & 0.60(0.14)  & 0.58(0.14)  \\
     \hline
     30-Minutes  &  0.59(0.23) & 0.67(0.21) & 0.69(0.22) \\  
     \hline
     60-Minutes  &  0.49(0.09) & 0.73(0.16)  & 0.72(0.14) \\
    \hline
     \hline
     \multicolumn{4}{|c|}{Healthy Study}\\
     \hline
     Window Length  & Precision & Accuracy & F1-Score  \\
     \hline
     10-Minutes  & 0.67 (0.21)   & 0.67 (0.18) & 0.45 (0.05)   \\
     \hline
     15-Minutes  & 0.64(0.23)  & 0.65 (0.19)  & 0.45(0.11)   \\
     \hline
     30-Minutes &0.73(0.10) & 0.63(0.09) & 0.78(0.14)\\
     \hline
     60-Minutes  & 0.69(0.24) & 0.71(0.18) & 0.46(0.03)  \\
    \hline

\end{tabular}
\caption{Receptivity Model Performance by Segment Window Length, using the Random Forest Algorithm.}
\label{tab:S3}
\end{table}


\begin{table}
    \centering

    \begin{tabular}{|c|c|c||c|c|}
    \hline
    Emotion Models& \multicolumn{2}{c||}{ADRD Study}& \multicolumn{2}{c|}{Healthy Study}\\
    \hline
     Algorithm & RMSE & $R^2$ & RMSE & $R^2$ \\
     \hline
    Neural Network& 4.47 (1.78) & -1.4(2.06)& 4.32 (1.13) & -0.90 (0.38)\\
     \hline
     Linear Regression & 4.8(1.9) & -3.3(5.0)& 5.30 (2.22)& -21.12 (18.97)\\
     \hline
     Baseline & 6.6(0.8) & -3.7(1.8) & 6.79 (1.29) & -1.11 (1.02)\\
     \hline
     
\end{tabular}
\caption{Emotion Model Performance Using LOSO Cross-Validation.}
\label{tab:S4}
\end{table}


\begin{table}
    \centering

    \begin{tabular}{|c|c|c|}
    \hline
     \multicolumn{3}{|c|}{ADRD Study}\\
    \hline
     Window Length  & RMSE & $R^2$  \\
     \hline
     10-Minutes  & 3.87 (0.04)  & 0.39 (0.1)     \\
     \hline
     15-Minutes  & 4.15 (0.04)  & 0.31 (0.12)     \\
     \hline
     30-Minutes  & 3.56 (0.02)  &  0.48 (0.01)    \\
     \hline
     60-Minutes  &  4.05 (0.07) &  0.12 (0.03)  \\
    \hline
     \hline
     \multicolumn{3}{|c|}{Healthy Study}\\
     \hline
     Window Length  & RMSE & $R^2$  \\
     \hline
     
     10-Minutes  & 4.4 (0.04) & 0.05 (0.02)    \\
     \hline
     15-Minutes  & 4.32 (0.02)  &  0.09 (0.01)  \\
     \hline
     30-Minutes  &4.15 (0.02) & 0.27 (0.1)\\
     \hline
     60-Minutes  & 4.30 (0.01) & 0.10 (0.01)  \\
    \hline

\end{tabular}
\caption{Emotion Model Performance by Segment Window Length, using Neural Network.}
\label{tab:S5}
\end{table}



\begin{table}
\centering  
\resizebox{0.95\linewidth}{!}{\begin{tabular}{p{0.1\linewidth} | p{0.3\linewidth} | p{0.5\linewidth}}
\hline
\textbf{Signal} &  \textbf{Features} & \textbf{Description}\\

\hline
\textbf{Steps}  & Mean, Median, Max, Standard Deviation, 75/25 percentile, maximum, Kurtosis, and Skew & Features are calculated using the total number of steps from each minute window across a five-minute and 30-minute interval. \\
\hline
\textbf{Heart Rate} & Mean, Median, Max, Min, Standard Deviation, 75/25 percentile, Interquartile Range, Root Mean Square, Kurtosis, and Skew  & Features are calculated using the heart rate values from each minute window across a five-minute and 30-minute interval. \\
\hline
\textbf{Heart Rate Variability (HRV)} & Mean, Median, Max, Standard Deviation, 75/25 percentile, Interquartile Range, and Root Mean Square & Features are calculated using the heart rate variability from the entire night. There are three HRV signals recorded Root Mean Square of Successive Differences between normal heartbeats (RMSSD), High Frequency (HF), and Low Frequency (LF). \\
\hline
\textbf{Sleep} & Time in Bed, Time Asleep, Sleep Efficiency, and Sleep Regularity Index  &  \\
\hline
\textbf{AWARE}  & Number of Text Messages Received, Number of Text Messages Sent, Number of Phone Calls, Number of missed calls, On-phone (Binary), Screen On (Locked vs. Unlocked), Location (Cluster Number) & AWARE data is collected periodically, log files for messages and calls contain the time of the event, type (received or made), and for calls the end time. Geolocation is sampled regularly and does not necessarily gather data when participants are in transit. Screen usage logs transition points, i.e., if the phone goes from locked to unlocked or vice versa.\\
\hline
\end{tabular}}
\caption{ADRD Dataset Features}
\label{tab:S6}
\end{table}

\begin{table}
\centering  
\resizebox{0.95\linewidth}{!}{\begin{tabular}{p{0.2\linewidth} | p{0.28\linewidth} | p{0.48\linewidth} }
\hline
\textbf{Signal} &  \textbf{Features} & \textbf{Description}\\
\hline
\textbf{Skin} \textbf{Temperature} \textbf{(ST)} & mean, median, mode, minimum, range, root mean square, zero cross, Kurtosis, skew, IQR 25\textsuperscript{th} and 75\textsuperscript{th} Percentile &Zero-cross here is based on the number of times ST crosses over the mean ST. Kurtosis measures the extremity of the data in the segment and skew is the measure of asymmetry.  \\
\hline
\textbf{Electro}\textbf{cardiogram} \textbf{(ECG)} & mean, median, mode, minimum, range, root mean square, zero cross Kurtosis, skew, IQR 25\textsuperscript{th} and 75\textsuperscript{th} Percentile RMSSD, CVSD, CVNNI SDNN, NNI50, NNI20, PNNI50, PNNI50, low frequency (lf), very low frequency (vlf), high frequency (hf), high/low frequency ratio (hf/lf) &  NN (N-N or R-R interval) indicates the time between heartbeats. NNI20/50 refers to the number of successive intervals that differ by more than 20 or 50 ms. P indicates the proportion of NNI20/50 in the segment. RMSSD is the root mean square of successive differences between heartbeats. CVNNI and CVSD are the coefficients of variation (sdnn/mean) and (rmssd/mean), respectively. Our frequency domain features are based on how much of the signal lies between 0.003 to 0.04 Hz (vlf),  0.04 to 0.15 Hz (lf), 0.15 to 0.40 Hz (hf) \\
\hline
\textbf{Electrodermal} \textbf{Activity} \textbf{(EDA)} & \textbf{Wavelet:} max, mean, std, median, above zero (1 second and half second wavelet)
\textbf{Raw:} amplitude, max, min, mean
\textbf{Filtered:} amplitude, max, min, average  & A 1-second and a half-second window were used for wavelet features. Features were calculated for both the first and second derivatives of each window size. 
\\
\hline
\end{tabular}}
\caption{Healthy Dataset Features}
\label{tab:S7}
\end{table}

\end{document}